\documentclass[prl,twocolumn,showpacs,superscriptaddress]{revtex4}

\usepackage{graphicx,epstopdf}
\usepackage{mathrsfs}
\usepackage{dcolumn}
\usepackage{bm}
\usepackage{amsmath}
\usepackage{amsfonts}
\usepackage{color}

\begin{document}

\title{Topological superconductivity on the surface of Fe-based superconductors}

\author{Gang Xu$^*$, Biao Lian\footnote{These two authors contributed equally to this work},
Peizhe Tang, Xiao-Liang Qi\footnote{xlqi@stanford.edu} and Shou-Cheng Zhang\footnote{sczhang@stanford.edu}}
\affiliation{Department of Physics, McCullough Building, Stanford University, Stanford, California 94305-4045, USA}

\date{\today}

\begin{abstract}
As one of the simplest systems for realizing Majorana fermions, topological superconductor plays an important role in both condensed matter physics and quantum computations. Based on \emph{ab~initio} calculations and the analysis of an effective 8-band model with the superconducting pairing, we demonstrate that the three dimensional extended $s$-wave Fe-based superconductors such as Fe$_{1+\text{y}}$Se$_{0.5}$Te$_{0.5}$ have a metallic topologically nontrivial band structure, and exhibit a normal-topological-normal superconductivity phase transition on the ($001$) surface by tuning the bulk carrier doping level. In the topological superconductivity (TSC) phase, a Majorana zero mode is trapped at the end of a magnetic vortex line. We further show that, the surface TSC phase only exists up to a certain bulk pairing gap, and there is a normal-topological phase transition driven by the temperature, which has not been discussed before. These results pave an effective way to realize the TSC and Majorana fermions in a large class of superconductors.
\end{abstract}
\pacs{74.20.-z, 74.70.Xa, 73.20.-r, 74.25.Uv}
\maketitle

TSC is known for its ability of hosting Majorana fermions and implementing topological quantum computations~\cite{Kitaev2001,Kitaev2003,Read2000}. As one of the most intriguing topics in today's physics research, lots of theoretical proposals have been raised for the realization of the Majorana zero modes (MZMs)~\cite{Sarma2006,nayak2008,fu2008,lee2009,qi2009b,qi2010,sau2010,lutchyn2010c,oreg2010,alicea2010, potter2011,Duckheim2011,Weng2011,Chung2011,Xu2014,Perge2014,Hao2014,Wu2014,Wang2015c}.
In particular, based on the topological insulator (TI), L. Fu and C. L. Kane proposed that a TSC can be achieved on the TI surface in proximity to the simplest $s$-wave superconductors, where the Dirac cone type surface states (SSs) are forced to favor a $p_x+ip_y$ pairing~\cite{fu2008}. To realize such a surface TSC, a lot of efforts have been devoted to the carrier doped TI such as Cu$_x$Bi$_2$Se$_3$~\cite{wray2010,Hor2010prl,Levy2013} and the  superconductor-TI heterostructures~\cite{Wang2013d,Yilmaz2014,Xu2014b}, in which some features suggesting the existence of MZMs have been observed but the direct evidences are still absent. However, there is little investigation along the other way of thinking, namely, looking for the intrinsic $s$-wave superconductors that possess a topologically nontrivial band structures and support the Dirac cone type SSs.

Recently, it is found by the density functional theory (DFT) calculations and confirmed by ARPES observations that superconducting (SC) FeSe$_{0.5}$Te$_{0.5}$ (FST) and LiFeAs possess topologically nontrivial band structures due to a band inversion in the $\Gamma-Z$ direction~\cite{Wang2015b}. At high temperatures, FST is a topologically nontrivial metal with a single Dirac cone on the surface. Below the SC transition temperature ($T_c = 14.5 K$)~\cite{Miao2012,Yin2015}, according to Fu and Kane's argument, the surface electrons in the Dirac cone have a chance to pair into a $p$-wave TSC due to the proximity effect of the bulk superconductivity. However, unlike Fu and Kane's model where the surface Dirac cone is in the TI band gap, the SSs in FST are buried in the metallic bulk bands. Therefore, whether the Cooper pairing of the SSs can form a TSC is an unknown question. In this letter, based on the DFT calculations and the analysis of an effective 8-band model with the SC pairing at the $\Gamma$ and $Z$ points, we clearly answer this question and predict a normal-topological-normal superconductivity phase transition on the ($001$) surface of a class of the extended s-wave superconductors such as FST, which possess topological nontrivial band structures around the Fermi level. In a proper chemical doping interval that can be easily achieved experimentally, the MZMs can be observed at the ends of a magnetic vortex line in FST. Compared to most previous proposals of TSC in heterostructures ~\cite{fu2008,lee2009,qi2009b,qi2010,sau2010,lutchyn2010c,oreg2010,alicea2010, potter2011,Duckheim2011,Weng2011,Chung2011,Xu2014,Perge2014}, TSC is realized within one material here, which at least leads to two advantages: 1) The complicated interactions and unpredictability at the interfaces are avoided, and the sample preparation and quality control becomes much easier in experiments. 2) A strong proximity effect between the bulk SC and the SSs is ensured. Our results suggest an efficient way to realize the TSC and Majorana fermions on the surface of such three-dimensional superconductors, which may have potential applications in the quantum computations.

\begin{figure}[tbp]
\includegraphics[clip,scale=0.43, angle=0]{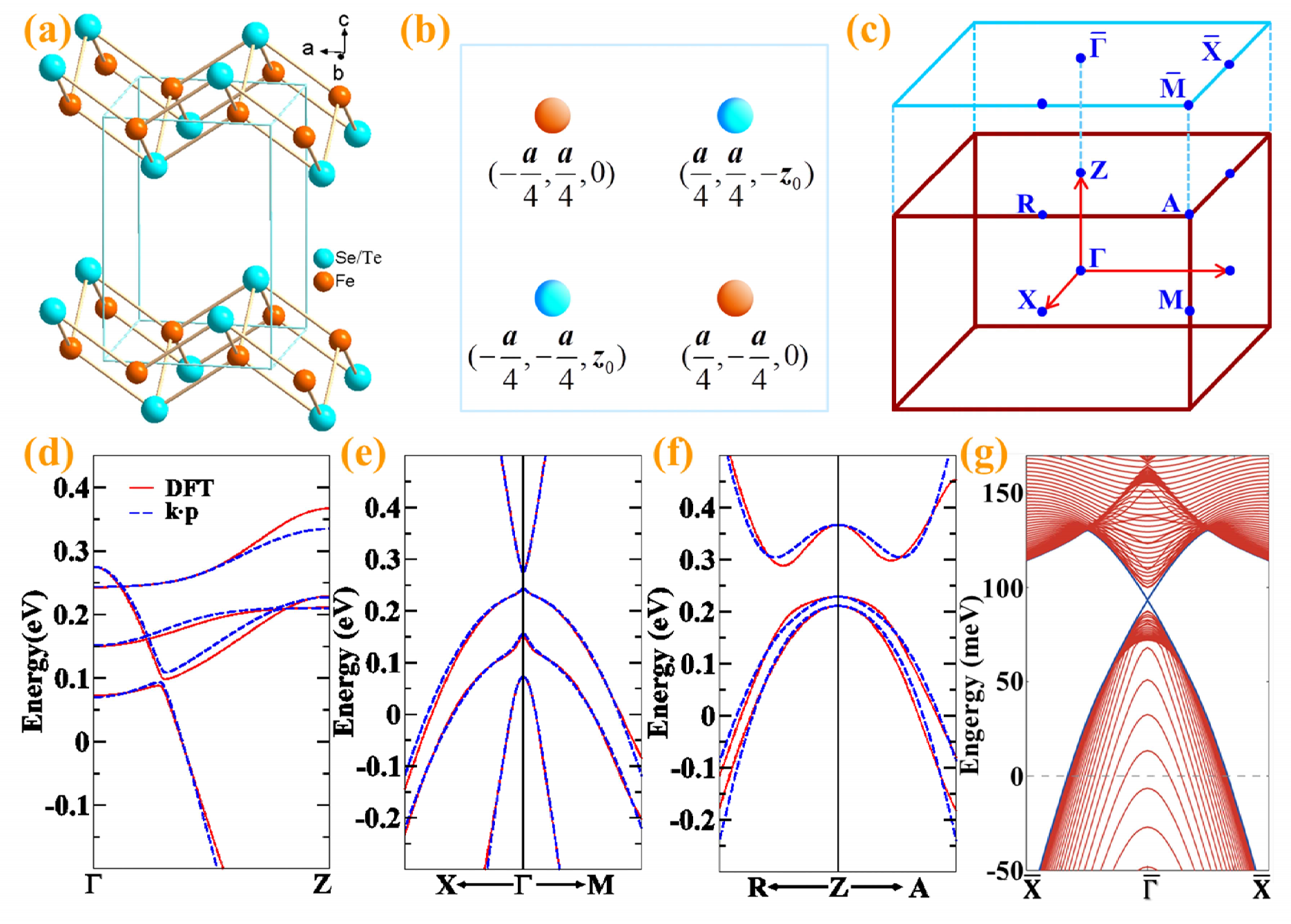}
\caption{(Color online) (a) Side view of the crystal structure of FST. (b) Top view of FST, where two Fe sites are marked by the brown circles, while two Se (Te) sites are marked by the light blue circles with $\pm z_0$ labeling their height from the Fe-plane. (c) The first BZ of FST with symmetry $P4/nmm$ and six types of inequivalent TRIPs. The light blue square shows the 2D BZ of the projected (001) surface, in which the high-symmetry k points $\bar{\Gamma}$, $\bar{X}$ and $\bar{M}$ are labeled. (d)-(f) The low energy dispersions of FST with SOC around the $\Gamma$ and $Z$ points. Red lines are the band structures calculated by the DFT calculations, while the blue dashes are the results from our effective model fitting. (g) The surface states calculated by the effective Hamiltonian and the parameters listed in Table I in the Supplemental Materials. In all figures, 0 eV corresponds to the Fermi level of the stoichiometric FST.\label{Fig1}}
\end{figure}

Experimentally, FST is synthesized within the inversion symmetric space group $P4/nmm$~\cite{Miao2012,Yin2015,fang2008,Li2009d,Bendele2010}, as shown in Fig. 1(a), where each layer of Fe atoms is sandwiched by two layers of Se (Te) atoms forming one unit cell and stacking along the $z$ direction. The first Brillouin zone (BZ) of such system is shown in Fig. 1(c), where there are eight time-reversal-invariant points (TRIPs), $\Gamma(0,0,0)$, $M(\pi,\pi,0)$, $Z(0,0,\pi)$, $A(\pi,\pi,\pi)$, two $R(\pi,0,\pi)$ and two $X(\pi,0,0)$. In the inversion symmetric system, the parity products of the TRIPs determine the $\mathbb{Z}_2$ topological index of the system~\cite{fu2007a,fu2007b}. Two equivalent $X$ points and two equivalent $R$ points always yield a trivial parity product. Besides, due to the negligible dispersion along the $M-A$ direction in the Fe-based superconductors, the parity product of the $M$ point and $A$ point is also trivial. Therefore, the parities of the occupied states at the $\Gamma$ and $Z$ points are the key to determine the topology of the electronic bands in FST (and any other Fe-based superconductors with $P4/nmm$ symmetry).

As shown in Fig. 1(b), each FST unit cell contains two Fe atoms that are quite close to each other (less than 2.7 \AA)~\cite{Li2009d}. The DFT calculations show that the $3d$-orbitals of Fe atoms dominate near the Fermi level~\cite{Wang2015b,xu2008a,xu2008b}. Under the appropriate consideration of the crystal symmetry, the bases describing the low energy bands near the $\Gamma$ and $Z$ points are simplified as Eq.S1 in the Supplemental Materials~\cite{SM}.
We note that the first three bases $|1\rangle$, $|2\rangle$ and $|3\rangle$ in Eq.S1 have an even parity, while the basis $|4\rangle$ has an odd parity. As we shall show below, the band inversion between bands $|2\rangle$ and $|4\rangle$ in the $\Gamma-Z$ direction leads to a topologically nontrivial band structure in FST.

The effective model at the $\Gamma$ point or $Z$ point has the full point group symmetry $D_{4h}$ of the crystal.
The full Hamiltonian with spin-orbit coupling (SOC) under the spinful bases $(|1\rangle,|2\rangle,|3\rangle,|4\rangle)\otimes(|\uparrow\rangle,|\downarrow\rangle)$ takes the form:
\begin{equation}
H(\mathbf{k})=H_0\otimes \mathbf{1}_2+H_{soc}
\end{equation}
where $H_0$ is a 4-band spinless Hamiltonian, and $H_{soc}$ is an 8$\times$8 matrix describing the SOC interaction. They are given explicitly in Eq.S2 and Eq.S4, respectively, in the Supplemental Materials~\cite{SM}.
The parameters of the effective Hamiltonian $H$ at the $\Gamma$ point and $Z$ point are listed in Table I in the Supplemental Materials, which are obtained by fitting with the DFT calculations~\cite{SM}. As shown in Fig. 1(d)-(f), the effective model (blue dashed lines) reproduces the band dispersions of the DFT calculations (red lines) well. In particular, the odd parity state $|4\rangle$ (the highest band at the $\Gamma$ point) is very dispersive along the $\Gamma-Z$ direction with a negative effective mass. As a result, it crosses with the other three even parity states in the $\Gamma-Z$ direction. In the presence of SOC, a topologically nontrivial band gap is opened between states $|2\rangle$ and $|4\rangle$ nearby the Fermi level (0 eV), while the crossing between state $|4\rangle$ and state $|1\rangle$ ($|3\rangle$) is protected by the crystalline symmetry. Fig. \ref{Fig1}(g) shows the energy spectrum of the effective Hamiltonian $H$ (using the parameters at the $\Gamma$ point) with an open boundary in the $z$ direction. Due to the nontrivial topology of the band structures, a surface Dirac cone arises, in consistency with the previous Green's function calculations~\cite{Wang2015b}.

When the superconductivity is considered, the electronic states on the FST ($001$) surface may fall into either a two dimensional (2D) normal superconductivity (NSC) phase or a TSC phase. It is usually believed that the surface is a TSC when the Fermi level crosses the surface Dirac cone, where the surface electrons occupy a single band and are thus forced to form a topologically nontrivial pairing under the bulk proximity effect~\cite{fu2008}. On the other hand, when the Fermi level is far away from the Dirac cone, all the electrons occupy the bulk bands and the surface is topologically trivial. Therefore, a surface phase transition from the NSC to the TSC is expected as the chemical potential approaches the Dirac cone. When a vortex line with $\pi$ magnetic flux is introduced in the bulk of the superconductor, there will be two MZMs (no MZM) trapped at the ends of the vortex line if the surface of the superconductor is TSC (NSC). For a better illustration, we have plotted the schematic evolution of the surface MZMs in the superconducting vortex line during the surface topological phase transition in Fig. 2(a)-(d): In the NSC phase, the vortex line is gapped and there is no surface MZMs (Fig. 2(a)). When the chemical potential is tuned to the transition point, the vortex line becomes gapless, as shown in Fig. 2(b). As the chemical potential is tuned into the TSC phase, the MZMs arise as shown in Fig. 2(c)-(d), whose localization length is inversely proportional to the bound state gap in the one dimensional (1D) bulk of the vortex line. This feature can be used to distinguish whether the surface of a three dimensional superconductor is TSC or not~\cite{Hosur2011}.

\begin{figure}[tbp]
\includegraphics[clip,scale=0.43, angle=0]{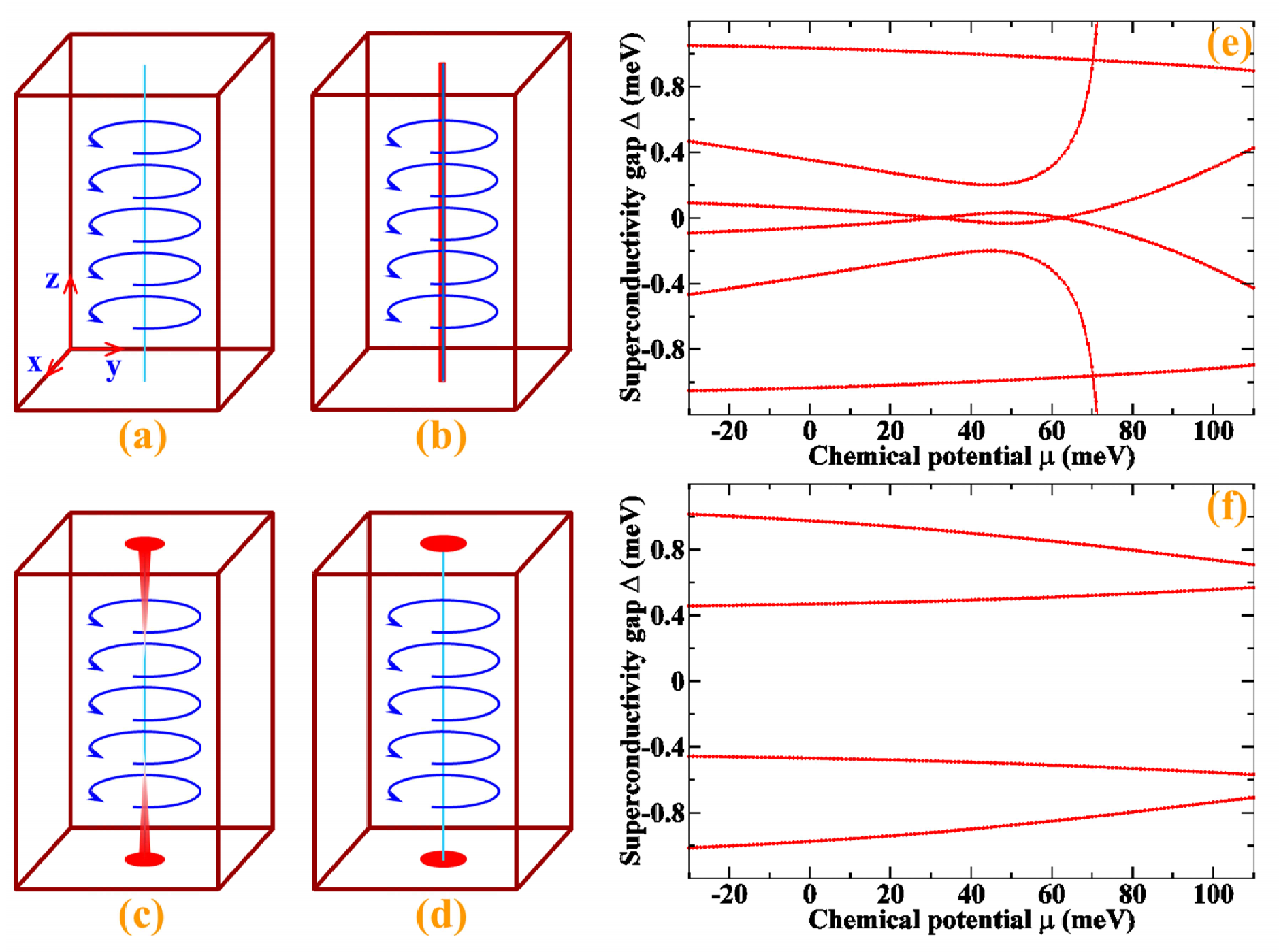}
\caption{(Color online) (a)-(d) The schematic evolution of the surface MZMs in a vortex line. As the chemical potential is tuned from the trivial regime (a) towards the surface TSC regime, two Majorana zero modes arise [(b)] and then become more and more localized at the ends of the vortex line [(c) and (d)]. (e) The energy spectrum at the $\Gamma$ point of a vortex line along the $z$ direction as a function of the chemical potential $\mu$. The energy gap closes at $\mu_1=31$meV and $\mu_2=62$meV, respectively, showing the (001) surface is a TSC in the range $\mu\in[\mu_1,\mu_2]$. (f) The energy spectrum at the $Z$ point of the vortex line as a function of the chemical potential $\mu$. There is no gap closing at the $Z$ point, so the phase transitions are solely determined by the gap closing at the $\Gamma$ point.
In all figures, chemical potential $\mu=0$ eV corresponds to the Fermi level of the stoichiometric FST.\label{Fig2}}
\end{figure}

\begin{figure}[tbp]
\includegraphics[clip,scale=0.43,angle=0]{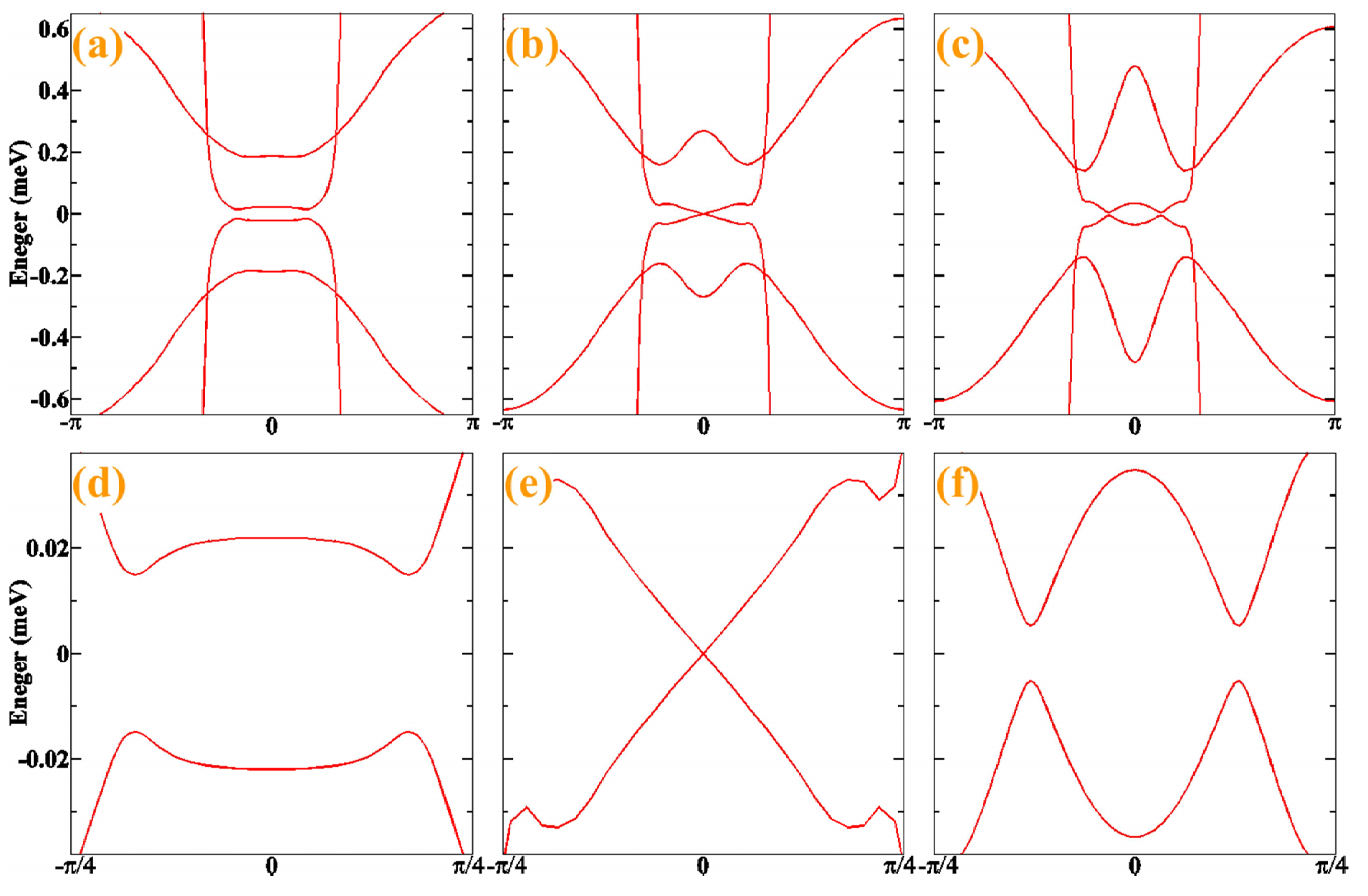}
\caption{(Color online) Low energy dispersions of the vortex line for the TSC phase [(a) and (d)], the transition point [(b) and (e)] and the NSC phase ([(c) and (f)], respectively. Here $\mu$ = 50 meV  and  $\mu$ = 70 meV are chosen to represent the TSC and NSC phase, while the results of the transition point are calculated at $\mu$ = 62 meV. (d), (e) and (f) are the enlargement of (a), (b) and (c) to show the gap clearly.~\label{Fig3}
}
\end{figure}

An equivalent understanding of the mechanism of the surface Majorana mode is to view the vortex line as a Majorana chain~\cite{Kitaev2001,Hosur2011}. Tuning the bulk Fermi level effectively varies the parameters of the Majorana chain, and thus drives a phase transition between 1D TSC and NSC. Such a phase transition is characterized by a gap closing of the BdG spectrum of the chain at high symmetry points, namely, either $k_z=0$ ($\Gamma$ point) or $k_z=\pi$ ($Z$ point)~\cite{Hosur2011}. Therefore, we numerically calculate the BdG spectrum on a vortex line along the $z$ direction in FST to determine the topological phase transition points. Since FST is an extreme type II superconductor with Ginzberg-Landau parameter $\kappa\approx180$~\cite{Lei2010,Kim2010}, the magnetic field in the vortex is extremely weak and can be ignored in the calculation. We take the following BdG Hamiltonian for the vortex line:
\begin{equation}
H_{BdG}=\left(\begin{array}{cc}H(\mathbf{k})-\mu&\Delta_{s} e^{i\theta}\tanh(r/\xi)\\ \Delta_{s} e^{-i\theta}\tanh(r/\xi)&-H(\mathbf{k})+\mu\end{array}\right)\
\end{equation}
where $H(\mathbf{k})$ is the effective 8-band Hamiltonian as shown explicitly in the Supplemental Materials~\cite{SM}, $\mu$ is the chemical potential, $(r,\theta)$ are the polar coordinates in the $\text{xy}$-plane. $\xi=3$ nm is the coherence length~\cite{Lei2010}, and $\Delta_s=\text{diag}(\Delta_{1},\Delta_{2},\Delta_{2},\Delta_{1})\otimes\mathbf{1}_2$ is the SC gap measured in the bulk FST, with $\Delta_{1}=2.5$meV and $\Delta_{2}=1.7$ meV describing the superconducting gap of the $d_{x^2-y^2}$ and $d_{xz}$ ($d_{yz}$) orbital, respectively~\cite{Miao2012,Yin2015}. Note that $k_z$ is still a good quantum number. By discretizing the polar coordinate $r$, we calculate the eigenvalues of the BdG Hamiltonian numerically at a given $\mu$ and $k_z$ on a disk with the radius $500$ nm. The calculated energy spectra at the $\Gamma$ point and $Z$ point are shown in Fig. \ref{Fig2} (e) and (f), respectively. The energy gap at the $\Gamma$ point closes at two chemical potentials $\mu_1=31$ meV and $\mu_2=62$ meV, while the spectrum at the $Z$ point is always gapped. Therefore, we expect TSC to be realized in the chemical potential interval $\mu\in[\mu_1,\mu_2]$. The energy dispersions on the vortex line are plotted at different chemical potentials in the TSC phase (Fig. \ref{Fig3} (a) and (d)), at the phase transition point (Fig. \ref{Fig3} (b), (e)) and in the trivial phase (Fig. \ref{Fig3} (c), (f)). As expected, the spectrum is always gapped away from the transition point, which allows a well-defined Zak phase $\theta_Z$ as the characteristic topological number~\cite{Hatsugai2006,Budich2013}. In particular, we calculate the Zak phases for $\mu$ = 50 meV and $\mu$ = 70 meV, respectively, and find that the Zak phase is $\theta_{Z} = \pi$ at $\mu$ = 50 meV, and is $\theta_{Z} = 0$ at $\mu$ = 70 meV. This verifies that the ($001$) surface of FST is a TSC in the chemical potential interval $\mu\in[\mu_1,\mu_2]$. We note that the energy interval $[\mu_1,\mu_2]$ is slightly lower than the energies of the surface Dirac cone shown in Fig. \ref{Fig1} (g), due to the particle-hole asymmetry of the bulk band structures~\cite{Chiu2012}. The transition points $\mu_1$ and $\mu_2$, however, does not agree with the $\pi$-Berry-phase criteria given in Ref.~\cite{Hosur2011} due to the multiple bands physics in FST (see Supplementary Material).

The MZMs at the ends of the vortex become more localized when the system is deeply in the TSC phase (Fig. \ref{Fig2} (c)-(d)). The localization length $l_M$ of the MZMs
can be estimated with the 1D bulk energy gap $E_0$ of the vortex line, which is up to $0.03$meV in the TSC phase as shown in Fig. \ref{Fig3}. With the bulk coherence length $\xi\approx\hbar v_F/\Delta_{1}\approx 3$ nm, we estimate the size of the Majorana zero mode to be $l_M\sim\hbar v_F/E_0= (\Delta_{1}/E_0)\xi\sim 10^2$ nm, where $v_F$ denotes the Fermi velocity in the bulk FST.

Our results show that one needs to dope some electrons into FST to realize a surface TSC. Fortunately, FST is such a material that it is usually self-doped by the excess Fe atoms when synthesized in experiments, as is denoted by the chemical formula Fe$_{1+\text{y}}$Se$_x$Te$_{1-x}$. In particular, the superconductivity of FST is robust in a wide range $-0.1<\text{y}<0.1$~\cite{fang2008,Li2009d,Bendele2010}.  To estimate the doping level for realizing the surface TSC, we perform the virtual crystal calculations for Fe$_{1+\text{y}}$Se$_{0.5}$Te$_{0.5}$~\cite{SM}, and plot the chemical potential $\mu$ as a function of the excess Fe content $\text{y}$ in Fig. S2~\cite{SM}. The chemical potential range for the surface TSC phase corresponds to $0.03<\text{y}<0.06$, which is well within the reach of the experiments. Besides, the chemical potential can also be tuned by an electrical gate voltage (2$\sim$3 V are needed). In a recent experiment work, using the scanning tunneling microscopy, Massee \emph{et al.}, observed a sharp zero bias peak at 0.25 K in the superconducting vortex core on the (001) surface of Fe(Te,Se)~\cite{Massee2015}, indicating the possible presence of the MZMs. Moreover,  according to our calculations discussed above, such a zero bias peak, if is induced by a MZM, should disappear when the chemical potential is tuned into the NSC regime via the excess Fe doping or electrical gating. Therefore, a further experimental verification of our predictions is necessary.

\begin{figure}[tbp]
\includegraphics[clip,scale=0.71,angle=0]{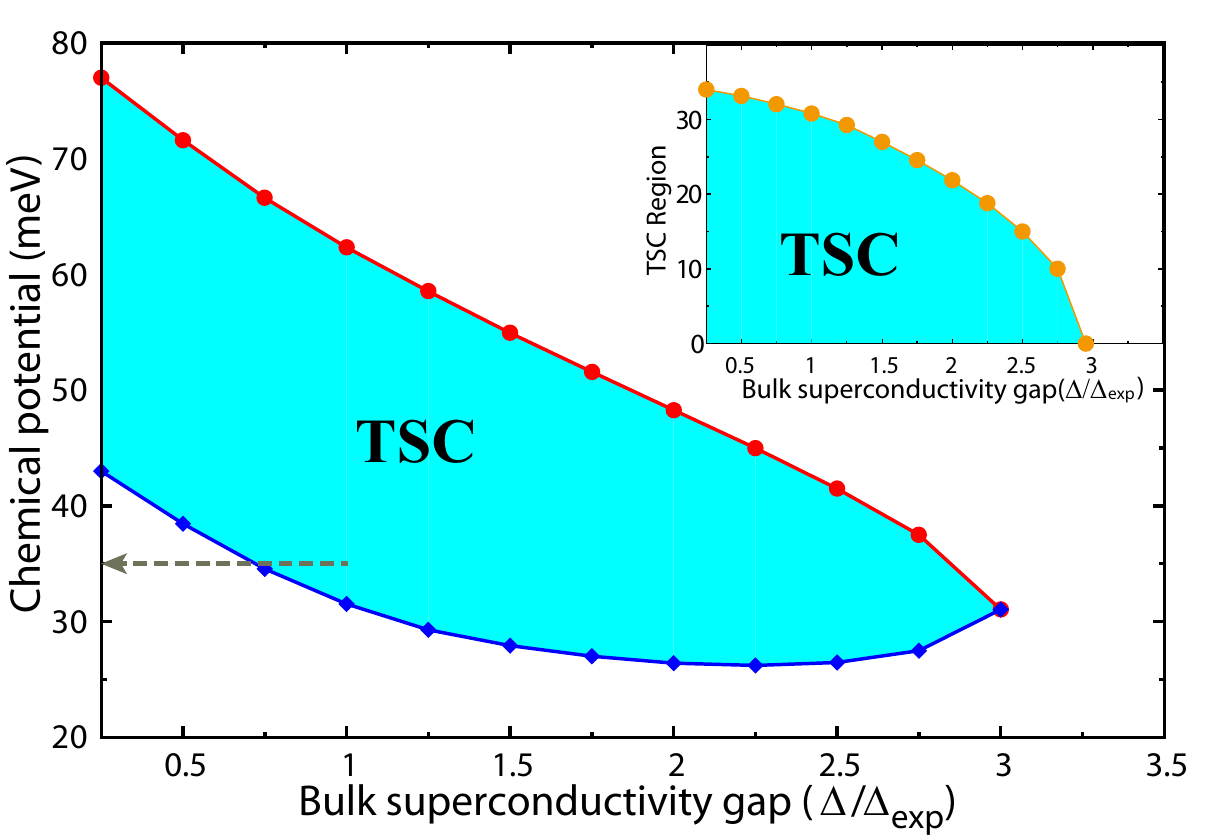}
\caption{(Color online) TSC phase space vs. bulk superconducting gap. The red line and blue line are the upper and lower phase boundaries of the TSC phase, respectively. The gray dash at $\mu = 35$ meV indicates a TSC at 0 $K$ ($\Delta$ = $\Delta_{exp}$) to NSC ($\Delta \sim 0$) phase transition with temperature increasing. The inset shows an evolution of the TSC region (red line minus blue line) with respect to the bulk pairing gap.~\label{Fig4}
}
\end{figure}

To investigate the dependence of the TSC regime on the bulk pairing gap $\Delta=(\Delta_1,\Delta_2)$, we vary $\Delta$ in the range up to $3.5\Delta_{exp}$, then calculate and plot the phase boundaries of the TSC as shown in Fig. \ref{Fig4}, where $\Delta_{exp}=(2.5 $meV$, 1.7$meV$)$ is the experimental pairing gap.
Our results show that the TSC phase is  narrowed and shifted towards the Fermi level as $\Delta$ increasing, and is bounded by a dome up to $\Delta=3\Delta_{exp}$ as shown in the inset of Fig. \ref{Fig4}. This is because that, when the bulk pairing increases and becomes more dominant, the SSs can pair with the bulk states more easily. Therefore, the surface TSC is suppressed and finally killed. Our result is theoretically a new discovery compared to the previous theories that only zero-pairing-gap limit is considered~\cite{fu2008,lee2009,qi2009b,Hosur2011}, where TSC phase always exists. Besides, due to the obvious particle-hole asymmetry of FST, the TSC dome is also particle-hole asymmetric.

In the real materials, the pairing gap $\Delta$ decreases as the temperature $T$ increases. Roughly, the BCS theory gives $\Delta(T)\sim\Delta_0\sqrt{1-T/T_c}$, where $\Delta_0$ is the zero temperature pairing gap. This enables one to see a phase transition from TSC to NSC by tuning the temperature $T$ of the system at a fixed chemical potentials. For instance, at $\mu=35$ meV of our results as shown in Fig. \ref{Fig4}, the surface of FST is in the TSC phase at $T=0$ ($\Delta \sim \Delta_{exp}$) and is trivial at $T=T_c$ ($\Delta \sim 0$). Ideally, one would expect there is a zero peak on the TSC side while a dip on the NSC side. However, the resolution of the zero peak will be reduced by the finite temperature, so that the transition might not be so sharp experimentally.

As a topologically protected phase, the surface TSC and the Majorana mode is robust against the weak disorders. To demonstrate this, we have introduced a random chemical potential disorder in the radial direction of the vortex line in our calculations. According to our calculations, up to $3\%$ impurity level as shown in the Fig. S3 of the Supplementary material~\cite{SM}, the disorder only shifts the TSC phase boundaries slightly.

Lastly, we note that such topologically nontrivial band structures are quite common in the Fe-based superconductors. Similar band inversion has been found in the other superconducting systems including LiFeAs (T$_c$=18 K) ~\cite{Wang2008,Wang2015b} and (Tl,Rb)$_\text{y}$Fe$_{2-x}$Se$_2$ (T$_c$=32 K)~\cite{Liu2012b}, which have the higher superconductivity transition temperatures and may support a larger-gap surface TSC. Accordingly, the physical mechanism of the surface TSC discussed in our article immediately applies for this large class of materials.

This work is supported by the U.S. Department of Energy, Office of Basic Energy Sciences, Division of Materials Sciences and Engineering under Contract No.~DE-AC02-76SF00515, by FAME, one of six centers of STARnet, a Semiconductor Research Corporation program sponsored by MARCO and DARPA, and supported by the NSF under grant numbers DMR-1305677. G. X. would also like to
thank for the support from NSF of China (11204359).


\begin{thebibliography}{48}
\expandafter\ifx\csname natexlab\endcsname\relax\def\natexlab#1{#1}\fi
\expandafter\ifx\csname bibnamefont\endcsname\relax
  \def\bibnamefont#1{#1}\fi
\expandafter\ifx\csname bibfnamefont\endcsname\relax
  \def\bibfnamefont#1{#1}\fi
\expandafter\ifx\csname citenamefont\endcsname\relax
  \def\citenamefont#1{#1}\fi
\expandafter\ifx\csname url\endcsname\relax
  \def\url#1{\texttt{#1}}\fi
\expandafter\ifx\csname urlprefix\endcsname\relax\def\urlprefix{URL }\fi
\providecommand{\bibinfo}[2]{#2}
\providecommand{\eprint}[2][]{\url{#2}}

\bibitem[{\citenamefont{Kitaev}(2001)}]{Kitaev2001}
\bibinfo{author}{\bibfnamefont{A.~Y.} \bibnamefont{Kitaev}},
  \bibinfo{journal}{Physics-Uspekhi} \textbf{\bibinfo{volume}{44}},
  \bibinfo{pages}{131} (\bibinfo{year}{2001}).

\bibitem[{\citenamefont{Kitaev}(2003)}]{Kitaev2003}
\bibinfo{author}{\bibfnamefont{A.~Y.} \bibnamefont{Kitaev}},
  \bibinfo{journal}{Annals of Physics} \textbf{\bibinfo{volume}{303}},
  \bibinfo{pages}{2 } (\bibinfo{year}{2003}).

\bibitem[{\citenamefont{Read and Green}(2000)}]{Read2000}
\bibinfo{author}{\bibfnamefont{N.}~\bibnamefont{Read}} \bibnamefont{and}
  \bibinfo{author}{\bibfnamefont{D.}~\bibnamefont{Green}},
  \bibinfo{journal}{Phys. Rev. B} \textbf{\bibinfo{volume}{61}},
  \bibinfo{pages}{10267} (\bibinfo{year}{2000}).

\bibitem[{\citenamefont{Das~Sarma et~al.}(2006)\citenamefont{Das~Sarma, Nayak,
  and Tewari}}]{Sarma2006}
\bibinfo{author}{\bibfnamefont{S.}~\bibnamefont{Das~Sarma}},
  \bibinfo{author}{\bibfnamefont{C.}~\bibnamefont{Nayak}}, \bibnamefont{and}
  \bibinfo{author}{\bibfnamefont{S.}~\bibnamefont{Tewari}},
  \bibinfo{journal}{Phys. Rev. B} \textbf{\bibinfo{volume}{73}},
  \bibinfo{pages}{220502} (\bibinfo{year}{2006}).

\bibitem[{\citenamefont{Nayak et~al.}(2008)\citenamefont{Nayak, Simon, Stern,
  Freedman, and Sarma}}]{nayak2008}
\bibinfo{author}{\bibfnamefont{C.}~\bibnamefont{Nayak}},
  \bibinfo{author}{\bibfnamefont{S.~H.} \bibnamefont{Simon}},
  \bibinfo{author}{\bibfnamefont{A.}~\bibnamefont{Stern}},
  \bibinfo{author}{\bibfnamefont{M.}~\bibnamefont{Freedman}}, \bibnamefont{and}
  \bibinfo{author}{\bibfnamefont{S.~D.} \bibnamefont{Sarma}},
  \bibinfo{journal}{Rev. Mod. Phys.} \textbf{\bibinfo{volume}{80}},
  \bibinfo{pages}{1083} (\bibinfo{year}{2008}).

\bibitem[{\citenamefont{Fu and Kane}(2008)}]{fu2008}
\bibinfo{author}{\bibfnamefont{L.}~\bibnamefont{Fu}} \bibnamefont{and}
  \bibinfo{author}{\bibfnamefont{C.~L.} \bibnamefont{Kane}},
  \bibinfo{journal}{Phys. Rev. Lett.} \textbf{\bibinfo{volume}{100}},
  \bibinfo{pages}{096407} (\bibinfo{year}{2008}).

\bibitem[{\citenamefont{Lee}(2009)}]{lee2009}
\bibinfo{author}{\bibfnamefont{P.~A.} \bibnamefont{Lee}},
  \bibinfo{journal}{arXiv:0907.2681}  (\bibinfo{year}{2009}).

\bibitem[{\citenamefont{Qi et~al.}(2009)\citenamefont{Qi, Hughes, Raghu, and
  Zhang}}]{qi2009b}
\bibinfo{author}{\bibfnamefont{X.-L.} \bibnamefont{Qi}},
  \bibinfo{author}{\bibfnamefont{T.~L.} \bibnamefont{Hughes}},
  \bibinfo{author}{\bibfnamefont{S.}~\bibnamefont{Raghu}}, \bibnamefont{and}
  \bibinfo{author}{\bibfnamefont{S.-C.} \bibnamefont{Zhang}},
  \bibinfo{journal}{Phys. Rev. Lett.} \textbf{\bibinfo{volume}{102}},
  \bibinfo{pages}{187001} (\bibinfo{year}{2009}).

\bibitem[{\citenamefont{Qi et~al.}(2010)\citenamefont{Qi, Hughes, and
  Zhang}}]{qi2010}
\bibinfo{author}{\bibfnamefont{X.-L.} \bibnamefont{Qi}},
  \bibinfo{author}{\bibfnamefont{T.~L.} \bibnamefont{Hughes}},
  \bibnamefont{and} \bibinfo{author}{\bibfnamefont{S.-C.} \bibnamefont{Zhang}},
  \bibinfo{journal}{Phys. Rev. B} \textbf{\bibinfo{volume}{82}},
  \bibinfo{pages}{184516} (\bibinfo{year}{2010}).

\bibitem[{\citenamefont{Sau et~al.}(2010)\citenamefont{Sau, Lutchyn, Tewari,
  and {Das Sarma}}}]{sau2010}
\bibinfo{author}{\bibfnamefont{J.~D.} \bibnamefont{Sau}},
  \bibinfo{author}{\bibfnamefont{R.~M.} \bibnamefont{Lutchyn}},
  \bibinfo{author}{\bibfnamefont{S.}~\bibnamefont{Tewari}}, \bibnamefont{and}
  \bibinfo{author}{\bibfnamefont{S.}~\bibnamefont{{Das Sarma}}},
  \bibinfo{journal}{Phys. Rev. Lett.} \textbf{\bibinfo{volume}{104}},
  \bibinfo{pages}{040502} (\bibinfo{year}{2010}).

\bibitem[{\citenamefont{Lutchyn et~al.}(2010)\citenamefont{Lutchyn, Sau, and
  Das~Sarma}}]{lutchyn2010c}
\bibinfo{author}{\bibfnamefont{R.~M.} \bibnamefont{Lutchyn}},
  \bibinfo{author}{\bibfnamefont{J.~D.} \bibnamefont{Sau}}, \bibnamefont{and}
  \bibinfo{author}{\bibfnamefont{S.}~\bibnamefont{Das~Sarma}},
  \bibinfo{journal}{Phys. Rev. Lett.} \textbf{\bibinfo{volume}{105}},
  \bibinfo{pages}{077001} (\bibinfo{year}{2010}).

\bibitem[{\citenamefont{Oreg et~al.}(2010)\citenamefont{Oreg, Refael, and von
  Oppen}}]{oreg2010}
\bibinfo{author}{\bibfnamefont{Y.}~\bibnamefont{Oreg}},
  \bibinfo{author}{\bibfnamefont{G.}~\bibnamefont{Refael}}, \bibnamefont{and}
  \bibinfo{author}{\bibfnamefont{F.}~\bibnamefont{von Oppen}},
  \bibinfo{journal}{Phys. Rev. Lett.} \textbf{\bibinfo{volume}{105}},
  \bibinfo{pages}{177002} (\bibinfo{year}{2010}).

\bibitem[{\citenamefont{Alicea}(2010)}]{alicea2010}
\bibinfo{author}{\bibfnamefont{J.}~\bibnamefont{Alicea}},
  \bibinfo{journal}{Phys. Rev. B} \textbf{\bibinfo{volume}{81}},
  \bibinfo{pages}{125318} (\bibinfo{year}{2010}).

\bibitem[{\citenamefont{Potter and Lee}(2011)}]{potter2011}
\bibinfo{author}{\bibfnamefont{A.~C.} \bibnamefont{Potter}} \bibnamefont{and}
  \bibinfo{author}{\bibfnamefont{P.~A.} \bibnamefont{Lee}},
  \bibinfo{journal}{Phys. Rev. B} \textbf{\bibinfo{volume}{83}},
  \bibinfo{pages}{094525} (\bibinfo{year}{2011}).

\bibitem[{\citenamefont{Duckheim and Brouwer}(2011)}]{Duckheim2011}
\bibinfo{author}{\bibfnamefont{M.}~\bibnamefont{Duckheim}} \bibnamefont{and}
  \bibinfo{author}{\bibfnamefont{P.~W.} \bibnamefont{Brouwer}},
  \bibinfo{journal}{Phys. Rev. B} \textbf{\bibinfo{volume}{83}},
  \bibinfo{pages}{054513} (\bibinfo{year}{2011}).

\bibitem[{\citenamefont{Weng et~al.}(2011)\citenamefont{Weng, Xu, Zhang, Zhang,
  Dai, and Fang}}]{Weng2011}
\bibinfo{author}{\bibfnamefont{H.}~\bibnamefont{Weng}},
  \bibinfo{author}{\bibfnamefont{G.}~\bibnamefont{Xu}},
  \bibinfo{author}{\bibfnamefont{H.}~\bibnamefont{Zhang}},
  \bibinfo{author}{\bibfnamefont{S.-C.} \bibnamefont{Zhang}},
  \bibinfo{author}{\bibfnamefont{X.}~\bibnamefont{Dai}}, \bibnamefont{and}
  \bibinfo{author}{\bibfnamefont{Z.}~\bibnamefont{Fang}},
  \bibinfo{journal}{Phys. Rev. B} \textbf{\bibinfo{volume}{84}},
  \bibinfo{pages}{060408} (\bibinfo{year}{2011}).

\bibitem[{\citenamefont{Chung et~al.}(2011)\citenamefont{Chung, Zhang, Qi, and
  Zhang}}]{Chung2011}
\bibinfo{author}{\bibfnamefont{S.~B.} \bibnamefont{Chung}},
  \bibinfo{author}{\bibfnamefont{H.-J.} \bibnamefont{Zhang}},
  \bibinfo{author}{\bibfnamefont{X.-L.} \bibnamefont{Qi}}, \bibnamefont{and}
  \bibinfo{author}{\bibfnamefont{S.-C.} \bibnamefont{Zhang}},
  \bibinfo{journal}{Phys. Rev. B} \textbf{\bibinfo{volume}{84}},
  \bibinfo{pages}{060510} (\bibinfo{year}{2011}).

\bibitem[{\citenamefont{Xu et~al.}(2014{\natexlab{a}})\citenamefont{Xu, Wang,
  Yan, and Qi}}]{Xu2014}
\bibinfo{author}{\bibfnamefont{G.}~\bibnamefont{Xu}},
  \bibinfo{author}{\bibfnamefont{J.}~\bibnamefont{Wang}},
  \bibinfo{author}{\bibfnamefont{B.}~\bibnamefont{Yan}}, \bibnamefont{and}
  \bibinfo{author}{\bibfnamefont{X.-L.} \bibnamefont{Qi}},
  \bibinfo{journal}{Phys. Rev. B} \textbf{\bibinfo{volume}{90}},
  \bibinfo{pages}{100505} (\bibinfo{year}{2014}{\natexlab{a}}).

\bibitem[{\citenamefont{Nadj-Perge et~al.}(2014)\citenamefont{Nadj-Perge,
  Drozdov, Li, Chen, Jeon, Seo, MacDonald, Bernevig, and Yazdani}}]{Perge2014}
\bibinfo{author}{\bibfnamefont{S.}~\bibnamefont{Nadj-Perge}},
  \bibinfo{author}{\bibfnamefont{I.~K.} \bibnamefont{Drozdov}},
  \bibinfo{author}{\bibfnamefont{J.}~\bibnamefont{Li}},
  \bibinfo{author}{\bibfnamefont{H.}~\bibnamefont{Chen}},
  \bibinfo{author}{\bibfnamefont{S.}~\bibnamefont{Jeon}},
  \bibinfo{author}{\bibfnamefont{J.}~\bibnamefont{Seo}},
  \bibinfo{author}{\bibfnamefont{A.~H.} \bibnamefont{MacDonald}},
  \bibinfo{author}{\bibfnamefont{B.~A.} \bibnamefont{Bernevig}},
  \bibnamefont{and} \bibinfo{author}{\bibfnamefont{A.}~\bibnamefont{Yazdani}},
  \bibinfo{journal}{Science} \textbf{\bibinfo{volume}{346}},
  \bibinfo{pages}{602} (\bibinfo{year}{2014}).

\bibitem[{\citenamefont{Hao and Hu}(2014)}]{Hao2014}
\bibinfo{author}{\bibfnamefont{N.}~\bibnamefont{Hao}} \bibnamefont{and}
  \bibinfo{author}{\bibfnamefont{J.}~\bibnamefont{Hu}}, \bibinfo{journal}{Phys.
  Rev. X} \textbf{\bibinfo{volume}{4}}, \bibinfo{pages}{031053}
  (\bibinfo{year}{2014}).

\bibitem[{\citenamefont{Wu et~al.}(2016)\citenamefont{Wu, Qin, Liang, Fan, and
  Hu}}]{Wu2014}
\bibinfo{author}{\bibfnamefont{X.}~\bibnamefont{Wu}},
  \bibinfo{author}{\bibfnamefont{S.}~\bibnamefont{Qin}},
  \bibinfo{author}{\bibfnamefont{Y.}~\bibnamefont{Liang}},
  \bibinfo{author}{\bibfnamefont{H.}~\bibnamefont{Fan}}, \bibnamefont{and}
  \bibinfo{author}{\bibfnamefont{J.}~\bibnamefont{Hu}}, \bibinfo{journal}{Phys.
  Rev. B} \textbf{\bibinfo{volume}{93}}, \bibinfo{pages}{115129}
  (\bibinfo{year}{2016}).

\bibitem[{\citenamefont{Wang and Liu}(2015)}]{Wang2015c}
\bibinfo{author}{\bibfnamefont{Q.-Z.} \bibnamefont{Wang}} \bibnamefont{and}
  \bibinfo{author}{\bibfnamefont{C.-X.} \bibnamefont{Liu}},
  \bibinfo{journal}{arXiv:1506.07938}  (\bibinfo{year}{2015}).

\bibitem[{\citenamefont{Wray et~al.}(2010)\citenamefont{Wray, Xu, Xia, Hor,
  Qian, Fedorov, Lin, Bansil, Cava, and Hasan}}]{wray2010}
\bibinfo{author}{\bibfnamefont{L.~A.} \bibnamefont{Wray}},
  \bibinfo{author}{\bibfnamefont{S.-Y.} \bibnamefont{Xu}},
  \bibinfo{author}{\bibfnamefont{Y.}~\bibnamefont{Xia}},
  \bibinfo{author}{\bibfnamefont{Y.~S.} \bibnamefont{Hor}},
  \bibinfo{author}{\bibfnamefont{D.}~\bibnamefont{Qian}},
  \bibinfo{author}{\bibfnamefont{A.~V.} \bibnamefont{Fedorov}},
  \bibinfo{author}{\bibfnamefont{H.}~\bibnamefont{Lin}},
  \bibinfo{author}{\bibfnamefont{A.}~\bibnamefont{Bansil}},
  \bibinfo{author}{\bibfnamefont{R.~J.} \bibnamefont{Cava}}, \bibnamefont{and}
  \bibinfo{author}{\bibfnamefont{M.~Z.} \bibnamefont{Hasan}},
  \bibinfo{journal}{Nat. Phys.} \textbf{\bibinfo{volume}{6}},
  \bibinfo{pages}{855} (\bibinfo{year}{2010}).

\bibitem[{\citenamefont{Hor et~al.}(2010)\citenamefont{Hor, Williams,
  Checkelsky, Roushan, Seo, Xu, Zandbergen, Yazdani, Ong, and
  Cava}}]{Hor2010prl}
\bibinfo{author}{\bibfnamefont{Y.~S.} \bibnamefont{Hor}},
  \bibinfo{author}{\bibfnamefont{A.~J.} \bibnamefont{Williams}},
  \bibinfo{author}{\bibfnamefont{J.~G.} \bibnamefont{Checkelsky}},
  \bibinfo{author}{\bibfnamefont{P.}~\bibnamefont{Roushan}},
  \bibinfo{author}{\bibfnamefont{J.}~\bibnamefont{Seo}},
  \bibinfo{author}{\bibfnamefont{Q.}~\bibnamefont{Xu}},
  \bibinfo{author}{\bibfnamefont{H.~W.} \bibnamefont{Zandbergen}},
  \bibinfo{author}{\bibfnamefont{A.}~\bibnamefont{Yazdani}},
  \bibinfo{author}{\bibfnamefont{N.~P.} \bibnamefont{Ong}}, \bibnamefont{and}
  \bibinfo{author}{\bibfnamefont{R.~J.} \bibnamefont{Cava}},
  \bibinfo{journal}{Phys. Rev. Lett.} \textbf{\bibinfo{volume}{104}},
  \bibinfo{pages}{057001} (\bibinfo{year}{2010}).

\bibitem[{\citenamefont{Levy et~al.}(2013)\citenamefont{Levy, Zhang, Ha,
  Sharifi, Talin, Kuk, and Stroscio}}]{Levy2013}
\bibinfo{author}{\bibfnamefont{N.}~\bibnamefont{Levy}},
  \bibinfo{author}{\bibfnamefont{T.}~\bibnamefont{Zhang}},
  \bibinfo{author}{\bibfnamefont{J.}~\bibnamefont{Ha}},
  \bibinfo{author}{\bibfnamefont{F.}~\bibnamefont{Sharifi}},
  \bibinfo{author}{\bibfnamefont{A.~A.} \bibnamefont{Talin}},
  \bibinfo{author}{\bibfnamefont{Y.}~\bibnamefont{Kuk}}, \bibnamefont{and}
  \bibinfo{author}{\bibfnamefont{J.~A.} \bibnamefont{Stroscio}},
  \bibinfo{journal}{Phys. Rev. Lett.} \textbf{\bibinfo{volume}{110}},
  \bibinfo{pages}{117001} (\bibinfo{year}{2013}).

\bibitem[{\citenamefont{Wang et~al.}(2013)\citenamefont{Wang, Ding, Fedorov,
  Yao, Li, Lv, Zhao, Zhang, Xu, Schneeloch et~al.}}]{Wang2013d}
\bibinfo{author}{\bibfnamefont{E.}~\bibnamefont{Wang}},
  \bibinfo{author}{\bibfnamefont{H.}~\bibnamefont{Ding}},
  \bibinfo{author}{\bibfnamefont{A.~V.} \bibnamefont{Fedorov}},
  \bibinfo{author}{\bibfnamefont{W.}~\bibnamefont{Yao}},
  \bibinfo{author}{\bibfnamefont{Z.}~\bibnamefont{Li}},
  \bibinfo{author}{\bibfnamefont{Y.-F.} \bibnamefont{Lv}},
  \bibinfo{author}{\bibfnamefont{K.}~\bibnamefont{Zhao}},
  \bibinfo{author}{\bibfnamefont{L.-G.} \bibnamefont{Zhang}},
  \bibinfo{author}{\bibfnamefont{Z.}~\bibnamefont{Xu}},
  \bibinfo{author}{\bibfnamefont{J.}~\bibnamefont{Schneeloch}},
  \bibnamefont{et~al.}, \bibinfo{journal}{Nature Phys.}
  \textbf{\bibinfo{volume}{9}}, \bibinfo{pages}{621} (\bibinfo{year}{2013}).

\bibitem[{\citenamefont{Yilmaz et~al.}(2014)\citenamefont{Yilmaz,
  Pletikosi\ifmmode~\acute{c}\else \'{c}\fi{}, Weber, Sadowski, Gu, Caruso,
  Sinkovic, and Valla}}]{Yilmaz2014}
\bibinfo{author}{\bibfnamefont{T.}~\bibnamefont{Yilmaz}},
  \bibinfo{author}{\bibfnamefont{I.}~\bibnamefont{Pletikosi\ifmmode~\acute{c}\else
  \'{c}\fi{}}}, \bibinfo{author}{\bibfnamefont{A.~P.} \bibnamefont{Weber}},
  \bibinfo{author}{\bibfnamefont{J.~T.} \bibnamefont{Sadowski}},
  \bibinfo{author}{\bibfnamefont{G.~D.} \bibnamefont{Gu}},
  \bibinfo{author}{\bibfnamefont{A.~N.} \bibnamefont{Caruso}},
  \bibinfo{author}{\bibfnamefont{B.}~\bibnamefont{Sinkovic}}, \bibnamefont{and}
  \bibinfo{author}{\bibfnamefont{T.}~\bibnamefont{Valla}},
  \bibinfo{journal}{Phys. Rev. Lett.} \textbf{\bibinfo{volume}{113}},
  \bibinfo{pages}{067003} (\bibinfo{year}{2014}).

\bibitem[{\citenamefont{Xu et~al.}(2014{\natexlab{b}})\citenamefont{Xu, Liu,
  Wang, Ge, Liu, Yang, Chen, Liu, Xu, Gao et~al.}}]{Xu2014b}
\bibinfo{author}{\bibfnamefont{J.-P.} \bibnamefont{Xu}},
  \bibinfo{author}{\bibfnamefont{C.}~\bibnamefont{Liu}},
  \bibinfo{author}{\bibfnamefont{M.-X.} \bibnamefont{Wang}},
  \bibinfo{author}{\bibfnamefont{J.}~\bibnamefont{Ge}},
  \bibinfo{author}{\bibfnamefont{Z.-L.} \bibnamefont{Liu}},
  \bibinfo{author}{\bibfnamefont{X.}~\bibnamefont{Yang}},
  \bibinfo{author}{\bibfnamefont{Y.}~\bibnamefont{Chen}},
  \bibinfo{author}{\bibfnamefont{Y.}~\bibnamefont{Liu}},
  \bibinfo{author}{\bibfnamefont{Z.-A.} \bibnamefont{Xu}},
  \bibinfo{author}{\bibfnamefont{C.-L.} \bibnamefont{Gao}},
  \bibnamefont{et~al.}, \bibinfo{journal}{Phys. Rev. Lett.}
  \textbf{\bibinfo{volume}{112}}, \bibinfo{pages}{217001}
  (\bibinfo{year}{2014}{\natexlab{b}}).

\bibitem[{\citenamefont{Wang et~al.}(2015)\citenamefont{Wang, Zhang, Xu, Zeng,
  Miao, Xu, Qian, Weng, Richard, Fedorov et~al.}}]{Wang2015b}
\bibinfo{author}{\bibfnamefont{Z.}~\bibnamefont{Wang}},
  \bibinfo{author}{\bibfnamefont{P.}~\bibnamefont{Zhang}},
  \bibinfo{author}{\bibfnamefont{G.}~\bibnamefont{Xu}},
  \bibinfo{author}{\bibfnamefont{L.~K.} \bibnamefont{Zeng}},
  \bibinfo{author}{\bibfnamefont{H.}~\bibnamefont{Miao}},
  \bibinfo{author}{\bibfnamefont{X.}~\bibnamefont{Xu}},
  \bibinfo{author}{\bibfnamefont{T.}~\bibnamefont{Qian}},
  \bibinfo{author}{\bibfnamefont{H.}~\bibnamefont{Weng}},
  \bibinfo{author}{\bibfnamefont{P.}~\bibnamefont{Richard}},
  \bibinfo{author}{\bibfnamefont{A.~V.} \bibnamefont{Fedorov}},
  \bibnamefont{et~al.}, \bibinfo{journal}{Phys. Rev. B}
  \textbf{\bibinfo{volume}{92}}, \bibinfo{pages}{115119}
  (\bibinfo{year}{2015}).

\bibitem[{\citenamefont{Miao et~al.}(2012)\citenamefont{Miao, Richard, Tanaka,
  Nakayama, Qian, Umezawa, Sato, Xu, Shi, Xu et~al.}}]{Miao2012}
\bibinfo{author}{\bibfnamefont{H.}~\bibnamefont{Miao}},
  \bibinfo{author}{\bibfnamefont{P.}~\bibnamefont{Richard}},
  \bibinfo{author}{\bibfnamefont{Y.}~\bibnamefont{Tanaka}},
  \bibinfo{author}{\bibfnamefont{K.}~\bibnamefont{Nakayama}},
  \bibinfo{author}{\bibfnamefont{T.}~\bibnamefont{Qian}},
  \bibinfo{author}{\bibfnamefont{K.}~\bibnamefont{Umezawa}},
  \bibinfo{author}{\bibfnamefont{T.}~\bibnamefont{Sato}},
  \bibinfo{author}{\bibfnamefont{Y.-M.} \bibnamefont{Xu}},
  \bibinfo{author}{\bibfnamefont{Y.~B.} \bibnamefont{Shi}},
  \bibinfo{author}{\bibfnamefont{N.}~\bibnamefont{Xu}}, \bibnamefont{et~al.},
  \bibinfo{journal}{Phys. Rev. B} \textbf{\bibinfo{volume}{85}},
  \bibinfo{pages}{094506} (\bibinfo{year}{2012}).

\bibitem[{\citenamefont{Yin et~al.}(2015)\citenamefont{Yin, Wu, Wang, Ye, Gong,
  Hou, Shan, Li, Liang, Wu et~al.}}]{Yin2015}
\bibinfo{author}{\bibfnamefont{J.-X.} \bibnamefont{Yin}},
  \bibinfo{author}{\bibfnamefont{Z.}~\bibnamefont{Wu}},
  \bibinfo{author}{\bibfnamefont{J.-H.} \bibnamefont{Wang}},
  \bibinfo{author}{\bibfnamefont{Z.-Y.} \bibnamefont{Ye}},
  \bibinfo{author}{\bibfnamefont{J.}~\bibnamefont{Gong}},
  \bibinfo{author}{\bibfnamefont{X.-Y.} \bibnamefont{Hou}},
  \bibinfo{author}{\bibfnamefont{L.}~\bibnamefont{Shan}},
  \bibinfo{author}{\bibfnamefont{A.}~\bibnamefont{Li}},
  \bibinfo{author}{\bibfnamefont{X.-J.} \bibnamefont{Liang}},
  \bibinfo{author}{\bibfnamefont{X.-X.} \bibnamefont{Wu}},
  \bibnamefont{et~al.}, \bibinfo{journal}{Nature Phys.}
  \textbf{\bibinfo{volume}{11}}, \bibinfo{pages}{543} (\bibinfo{year}{2015}).

\bibitem[{\citenamefont{Fang et~al.}(2008)\citenamefont{Fang, Pham, Qian, Liu,
  Vehstedt, Liu, Spinu, and Mao}}]{fang2008}
\bibinfo{author}{\bibfnamefont{M.~H.} \bibnamefont{Fang}},
  \bibinfo{author}{\bibfnamefont{H.~M.} \bibnamefont{Pham}},
  \bibinfo{author}{\bibfnamefont{B.}~\bibnamefont{Qian}},
  \bibinfo{author}{\bibfnamefont{T.~J.} \bibnamefont{Liu}},
  \bibinfo{author}{\bibfnamefont{E.~K.} \bibnamefont{Vehstedt}},
  \bibinfo{author}{\bibfnamefont{Y.}~\bibnamefont{Liu}},
  \bibinfo{author}{\bibfnamefont{L.}~\bibnamefont{Spinu}}, \bibnamefont{and}
  \bibinfo{author}{\bibfnamefont{Z.~Q.} \bibnamefont{Mao}},
  \bibinfo{journal}{Phys. Rev. B} \textbf{\bibinfo{volume}{78}},
  \bibinfo{pages}{224503} (\bibinfo{year}{2008}).

\bibitem[{\citenamefont{Li et~al.}(2009)\citenamefont{Li, de~la Cruz, Huang,
  Chen, Lynn, Hu, Huang, Hsu, Yeh, Wu et~al.}}]{Li2009d}
\bibinfo{author}{\bibfnamefont{S.}~\bibnamefont{Li}},
  \bibinfo{author}{\bibfnamefont{C.}~\bibnamefont{de~la Cruz}},
  \bibinfo{author}{\bibfnamefont{Q.}~\bibnamefont{Huang}},
  \bibinfo{author}{\bibfnamefont{Y.}~\bibnamefont{Chen}},
  \bibinfo{author}{\bibfnamefont{J.~W.} \bibnamefont{Lynn}},
  \bibinfo{author}{\bibfnamefont{J.}~\bibnamefont{Hu}},
  \bibinfo{author}{\bibfnamefont{Y.-L.} \bibnamefont{Huang}},
  \bibinfo{author}{\bibfnamefont{F.-C.} \bibnamefont{Hsu}},
  \bibinfo{author}{\bibfnamefont{K.-W.} \bibnamefont{Yeh}},
  \bibinfo{author}{\bibfnamefont{M.-K.} \bibnamefont{Wu}},
  \bibnamefont{et~al.}, \bibinfo{journal}{Phys. Rev. B}
  \textbf{\bibinfo{volume}{79}}, \bibinfo{pages}{054503}
  (\bibinfo{year}{2009}).

\bibitem[{\citenamefont{Bendele et~al.}(2010)\citenamefont{Bendele, Babkevich,
  Katrych, Gvasaliya, Pomjakushina, Conder, Roessli, Boothroyd, Khasanov, and
  Keller}}]{Bendele2010}
\bibinfo{author}{\bibfnamefont{M.}~\bibnamefont{Bendele}},
  \bibinfo{author}{\bibfnamefont{P.}~\bibnamefont{Babkevich}},
  \bibinfo{author}{\bibfnamefont{S.}~\bibnamefont{Katrych}},
  \bibinfo{author}{\bibfnamefont{S.~N.} \bibnamefont{Gvasaliya}},
  \bibinfo{author}{\bibfnamefont{E.}~\bibnamefont{Pomjakushina}},
  \bibinfo{author}{\bibfnamefont{K.}~\bibnamefont{Conder}},
  \bibinfo{author}{\bibfnamefont{B.}~\bibnamefont{Roessli}},
  \bibinfo{author}{\bibfnamefont{A.~T.} \bibnamefont{Boothroyd}},
  \bibinfo{author}{\bibfnamefont{R.}~\bibnamefont{Khasanov}}, \bibnamefont{and}
  \bibinfo{author}{\bibfnamefont{H.}~\bibnamefont{Keller}},
  \bibinfo{journal}{Phys. Rev. B} \textbf{\bibinfo{volume}{82}},
  \bibinfo{pages}{212504} (\bibinfo{year}{2010}).

\bibitem[{\citenamefont{Fu and Kane}(2007)}]{fu2007a}
\bibinfo{author}{\bibfnamefont{L.}~\bibnamefont{Fu}} \bibnamefont{and}
  \bibinfo{author}{\bibfnamefont{C.~L.} \bibnamefont{Kane}},
  \bibinfo{journal}{Phys. Rev. B} \textbf{\bibinfo{volume}{76}},
  \bibinfo{pages}{045302} (\bibinfo{year}{2007}).

\bibitem[{\citenamefont{Fu et~al.}(2007)\citenamefont{Fu, Kane, and
  Mele}}]{fu2007b}
\bibinfo{author}{\bibfnamefont{L.}~\bibnamefont{Fu}},
  \bibinfo{author}{\bibfnamefont{C.~L.} \bibnamefont{Kane}}, \bibnamefont{and}
  \bibinfo{author}{\bibfnamefont{E.~J.} \bibnamefont{Mele}},
  \bibinfo{journal}{Phys. Rev. Lett.} \textbf{\bibinfo{volume}{98}},
  \bibinfo{pages}{106803} (\bibinfo{year}{2007}).

\bibitem[{\citenamefont{Xu et~al.}(2008{\natexlab{a}})\citenamefont{Xu, Ming,
  Yao, Dai, Zhang, and Fang}}]{xu2008a}
\bibinfo{author}{\bibfnamefont{G.}~\bibnamefont{Xu}},
  \bibinfo{author}{\bibfnamefont{W.}~\bibnamefont{Ming}},
  \bibinfo{author}{\bibfnamefont{Y.}~\bibnamefont{Yao}},
  \bibinfo{author}{\bibfnamefont{X.}~\bibnamefont{Dai}},
  \bibinfo{author}{\bibfnamefont{S.-C.} \bibnamefont{Zhang}}, \bibnamefont{and}
  \bibinfo{author}{\bibfnamefont{Z.}~\bibnamefont{Fang}},
  \bibinfo{journal}{Europhysics Letters} \textbf{\bibinfo{volume}{82}},
  \bibinfo{pages}{67002} (\bibinfo{year}{2008}{\natexlab{a}}).

\bibitem[{\citenamefont{Xu et~al.}(2008{\natexlab{b}})\citenamefont{Xu, Zhang,
  Dai, and Fang}}]{xu2008b}
\bibinfo{author}{\bibfnamefont{G.}~\bibnamefont{Xu}},
  \bibinfo{author}{\bibfnamefont{H.}~\bibnamefont{Zhang}},
  \bibinfo{author}{\bibfnamefont{X.}~\bibnamefont{Dai}}, \bibnamefont{and}
  \bibinfo{author}{\bibfnamefont{Z.}~\bibnamefont{Fang}},
  \bibinfo{journal}{Europhysics Letters} \textbf{\bibinfo{volume}{84}},
  \bibinfo{pages}{67015} (\bibinfo{year}{2008}{\natexlab{b}}).

\bibitem{SM} See Supplemental Materials for more details of the DFT calculations, the effective Hamiltonian and parameters,
SU(2) Berry phase calculation and the disorder calculations.

\bibitem[{\citenamefont{Hosur et~al.}(2011)\citenamefont{Hosur, Ghaemi, Mong,
  and Vishwanath}}]{Hosur2011}
\bibinfo{author}{\bibfnamefont{P.}~\bibnamefont{Hosur}},
  \bibinfo{author}{\bibfnamefont{P.}~\bibnamefont{Ghaemi}},
  \bibinfo{author}{\bibfnamefont{R.~S.~K.} \bibnamefont{Mong}},
  \bibnamefont{and}
  \bibinfo{author}{\bibfnamefont{A.}~\bibnamefont{Vishwanath}},
  \bibinfo{journal}{Phys. Rev. Lett.} \textbf{\bibinfo{volume}{107}},
  \bibinfo{pages}{097001} (\bibinfo{year}{2011}).

\bibitem[{\citenamefont{Lei et~al.}(2010)\citenamefont{Lei, Hu, Choi, Warren,
  and Petrovic}}]{Lei2010}
\bibinfo{author}{\bibfnamefont{H.}~\bibnamefont{Lei}},
  \bibinfo{author}{\bibfnamefont{R.}~\bibnamefont{Hu}},
  \bibinfo{author}{\bibfnamefont{E.~S.} \bibnamefont{Choi}},
  \bibinfo{author}{\bibfnamefont{J.~B.} \bibnamefont{Warren}},
  \bibnamefont{and} \bibinfo{author}{\bibfnamefont{C.}~\bibnamefont{Petrovic}},
  \bibinfo{journal}{Phys. Rev. B} \textbf{\bibinfo{volume}{81}},
  \bibinfo{pages}{094518} (\bibinfo{year}{2010}).

\bibitem[{\citenamefont{Kim et~al.}(2010)\citenamefont{Kim, Martin, Gordon,
  Tanatar, Hu, Qian, Mao, Hu, Petrovic, Salovich et~al.}}]{Kim2010}
\bibinfo{author}{\bibfnamefont{H.}~\bibnamefont{Kim}},
  \bibinfo{author}{\bibfnamefont{C.}~\bibnamefont{Martin}},
  \bibinfo{author}{\bibfnamefont{R.~T.} \bibnamefont{Gordon}},
  \bibinfo{author}{\bibfnamefont{M.~A.} \bibnamefont{Tanatar}},
  \bibinfo{author}{\bibfnamefont{J.}~\bibnamefont{Hu}},
  \bibinfo{author}{\bibfnamefont{B.}~\bibnamefont{Qian}},
  \bibinfo{author}{\bibfnamefont{Z.~Q.} \bibnamefont{Mao}},
  \bibinfo{author}{\bibfnamefont{R.}~\bibnamefont{Hu}},
  \bibinfo{author}{\bibfnamefont{C.}~\bibnamefont{Petrovic}},
  \bibinfo{author}{\bibfnamefont{N.}~\bibnamefont{Salovich}},
  \bibnamefont{et~al.}, \bibinfo{journal}{Phys. Rev. B}
  \textbf{\bibinfo{volume}{81}}, \bibinfo{pages}{180503}
  (\bibinfo{year}{2010}).

\bibitem[{\citenamefont{Hatsugai}(2006)}]{Hatsugai2006}
\bibinfo{author}{\bibfnamefont{Y.}~\bibnamefont{Hatsugai}},
  \bibinfo{journal}{Journal of the Physical Society of Japan}
  \textbf{\bibinfo{volume}{75}}, \bibinfo{pages}{123601}
  (\bibinfo{year}{2006}).

\bibitem[{\citenamefont{Budich and Ardonne}(2013)}]{Budich2013}
\bibinfo{author}{\bibfnamefont{J.~C.} \bibnamefont{Budich}} \bibnamefont{and}
  \bibinfo{author}{\bibfnamefont{E.}~\bibnamefont{Ardonne}},
  \bibinfo{journal}{Phys. Rev. B} \textbf{\bibinfo{volume}{88}},
  \bibinfo{pages}{075419} (\bibinfo{year}{2013}).

\bibitem[{\citenamefont{Chiu et~al.}(2012)\citenamefont{Chiu, Ghaemi, and
  Hughes}}]{Chiu2012}
\bibinfo{author}{\bibfnamefont{C.-K.} \bibnamefont{Chiu}},
  \bibinfo{author}{\bibfnamefont{P.}~\bibnamefont{Ghaemi}}, \bibnamefont{and}
  \bibinfo{author}{\bibfnamefont{T.~L.} \bibnamefont{Hughes}},
  \bibinfo{journal}{Phys. Rev. Lett.} \textbf{\bibinfo{volume}{109}},
  \bibinfo{pages}{237009} (\bibinfo{year}{2012}).

\bibitem[{\citenamefont{Massee et~al.}(2015)\citenamefont{Massee, Sprau, Wang,
  Davis, Ghigo, Gu, and Kwok}}]{Massee2015}
\bibinfo{author}{\bibfnamefont{F.}~\bibnamefont{Massee}},
  \bibinfo{author}{\bibfnamefont{P.~O.} \bibnamefont{Sprau}},
  \bibinfo{author}{\bibfnamefont{Y.-L.} \bibnamefont{Wang}},
  \bibinfo{author}{\bibfnamefont{J.~C.~S.} \bibnamefont{Davis}},
  \bibinfo{author}{\bibfnamefont{G.}~\bibnamefont{Ghigo}},
  \bibinfo{author}{\bibfnamefont{G.~D.} \bibnamefont{Gu}}, \bibnamefont{and}
  \bibinfo{author}{\bibfnamefont{W.-K.} \bibnamefont{Kwok}},
  \bibinfo{journal}{Science Advances} \textbf{\bibinfo{volume}{1}},
  \bibinfo{pages}{e1500033} (\bibinfo{year}{2015}).

\bibitem[{\citenamefont{Wang et~al.}(2008)\citenamefont{Wang, Liu, Lv, Gao,
  Yang, Yu, Li, and Jin}}]{Wang2008}
\bibinfo{author}{\bibfnamefont{X.}~\bibnamefont{Wang}},
  \bibinfo{author}{\bibfnamefont{Q.}~\bibnamefont{Liu}},
  \bibinfo{author}{\bibfnamefont{Y.}~\bibnamefont{Lv}},
  \bibinfo{author}{\bibfnamefont{W.}~\bibnamefont{Gao}},
  \bibinfo{author}{\bibfnamefont{L.}~\bibnamefont{Yang}},
  \bibinfo{author}{\bibfnamefont{R.}~\bibnamefont{Yu}},
  \bibinfo{author}{\bibfnamefont{F.}~\bibnamefont{Li}}, \bibnamefont{and}
  \bibinfo{author}{\bibfnamefont{C.}~\bibnamefont{Jin}},
  \bibinfo{journal}{Solid State Communications} \textbf{\bibinfo{volume}{148}},
  \bibinfo{pages}{538 } (\bibinfo{year}{2008}).

\bibitem[{\citenamefont{Liu et~al.}(2012)\citenamefont{Liu, Richard, Xu, Xu,
  Li, Fang, Jia, Chen, Wang, He et~al.}}]{Liu2012b}
\bibinfo{author}{\bibfnamefont{Z.-H.} \bibnamefont{Liu}},
  \bibinfo{author}{\bibfnamefont{P.}~\bibnamefont{Richard}},
  \bibinfo{author}{\bibfnamefont{N.}~\bibnamefont{Xu}},
  \bibinfo{author}{\bibfnamefont{G.}~\bibnamefont{Xu}},
  \bibinfo{author}{\bibfnamefont{Y.}~\bibnamefont{Li}},
  \bibinfo{author}{\bibfnamefont{X.-C.} \bibnamefont{Fang}},
  \bibinfo{author}{\bibfnamefont{L.-L.} \bibnamefont{Jia}},
  \bibinfo{author}{\bibfnamefont{G.-F.} \bibnamefont{Chen}},
  \bibinfo{author}{\bibfnamefont{D.-M.} \bibnamefont{Wang}},
  \bibinfo{author}{\bibfnamefont{J.-B.} \bibnamefont{He}},
  \bibnamefont{et~al.}, \bibinfo{journal}{Phys. Rev. Lett.}
  \textbf{\bibinfo{volume}{109}}, \bibinfo{pages}{037003}
  (\bibinfo{year}{2012}).

\end{thebibliography}
\end{document}


\renewcommand{\thefigure}{S\arabic{figure}}

\title{Supplemental Materials for 'Topological superconductivity on the surface of Fe-based superconductors'}

\author{Gang Xu$^*$, Biao Lian\footnote{These two authors contributed equally to this work},
Peizhe Tang, Xiao-Liang Qi\footnote{xlqi@stanford.edu} and Shou-Cheng Zhang\footnote{sczhang@stanford.edu}}

\affiliation{Department of Physics, McCullough Building, Stanford University, Stanford, California 94305-4045, USA}

\date{\today}

\pacs{74.20.-z, 74.70.Xa, 73.20.-r, 74.25.Uv}
\maketitle

\section{DFT calculations}

Our density functional theory (DFT) calculations are carried out by the BSTATE (Beijing Simulation Tool of Atomic Technology) package~\cite{Fang2002} based on the density functional theory~\cite{Hohenberg1964,Kohn1965} within the plane-wave ultrasoft pseudopotential scheme~\cite{Vanderbilt1990}. We use the generalized gradient approximation (GGA) of the Perdew-Burke-Ernzerhof type for the exchange-correlation potential~\cite{Perdew1996}. The experimental lattice parameters and internal coordinate with $a=3.7933 \AA$, $c=5.9552 \AA$ and $z_0 =0.2719$~\cite{Li2009d} are used in the calculations, and the virtual crystal approximation~\cite{Nordheim1931,Bellaiche2000} is used to simulate the Te substitution. The cutoff energy for the wave function expansion is set to 500 eV, and $20\times20\times12$ k meshes are used for the self-consistent calculations. Spin-orbit coupling (SOC) effect is consistently considered in all calculations.

\section{Effective model and parameters}

As shown in Fig. 1(a) and 1(b) in the main text, each FeSe$_{0.5}$Te$_{0.5}$ (FST) unit cell contains two Fe atoms, which are very close to each other (less than 2.7 \AA)~\cite{Li2009d}. So the Fe-Fe bonding is important to obtain the right band dispersions and orbital components. Therefore, considering our DFT calculations and previous studies~\cite{Wang2015b,xu2008a,xu2008b}, we choose the following basis to describe the low energy band structures around the $\Gamma$ and Z point:

\begin{equation} \tag{Eq.S1}\label{basis}
\begin{split}
&|1\rangle=\frac{1}{\sqrt{2}}\left(\phi^A_{x^2-y^2}+\phi^B_{x^2-y^2}\right)\ , \\
&|2\rangle=\frac{1}{2}\left[\left(\phi^A_{yz}+\phi^B_{yz}\right)+i\left(\phi^A_{xz}+\phi^B_{xz}\right)\right]\ , \\
&|3\rangle=\frac{1}{2}\left[\left(\phi^A_{yz}+\phi^B_{yz}\right)-i\left(\phi^A_{xz}+\phi^B_{xz}\right)\right]\ , \\
&|4\rangle=\frac{1}{\sqrt{2}}\left(\phi^A_{x^2-y^2}-\phi^B_{x^2-y^2}\right)\ .
\end{split}
\end{equation}
where $\phi^A$ and $\phi^B$ are the $3d$-orbitals of the Fe atoms at site $A(-\frac{a}{4},\frac{a}{4},0)$ and site $B(\frac{a}{4},-\frac{a}{4},0)$, respectively, as shown in Fig. 1(b) in the main text. Here, the first three bases $|1\rangle$, $|2\rangle$ and $|3\rangle$ have an even parity, while the basis $|4\rangle$ has an odd parity. As we discussed in the main text, the band inversion between bands $|2\rangle$ and $|4\rangle$ in the $\Gamma-Z$ direction leads to a topologically nontrivial band structures in FST.

The effective model at the $\Gamma$ point or $Z$ point has the full point group symmetry $D_{4h}$ of the crystal, whose generators include Inversion $I$, Mirror of x=0 plane $M_{x}$, and Rotation around z axis $R_{4z}$. By symmetry analysis, the 4-band effective model $H_0(\mathbf{k})$ under the basis Eq. S1 without SOC is restricted to the following form:
\begin{widetext}
\begin{equation}\tag{Eq.S2} \label{H0}
H_0(\mathbf{k})=\left(\begin{array}{cccc}M_1(\mathbf{k})&\gamma\sin(ck_z)(k_x-ik_y)&\gamma\sin(ck_z)(k_x+ik_y)&0 \\ \gamma\sin(ck_z)(k_x+ik_y)& M_2(\mathbf{k}) & \beta(k_y^2-k_x^2)+i\alpha k_xk_y & i\delta(k_x-ik_y)\\ \gamma\sin(ck_z)(k_x-ik_y)& \beta(k_y^2-k_x^2)-i\alpha k_xk_y & M_2(\mathbf{k}) & i\delta(k_x+ik_y)\\ 0& -i\delta(k_x+ik_y) & -i\delta(k_x-ik_y) & M_4(\mathbf{k})\\ \end{array}\right)
\end{equation}
\end{widetext}
where we have defined

\begin{equation}\tag{Eq.S3} \label{Mk}
M_n(\mathbf{k})=E_n+\frac{k_x^2+k_y^2}{2m_{nx}}+t_{nz}({1-\cos(ck_z)}) ~~~ n=1,2,4\ .
\end{equation}$E_n$ is the energy of the band $n$ at the $\Gamma$ ($Z$) point, while $m_{nx}$ and $t_{nz}$ are the in-plane effective mass and the $z$ direction hopping of the band $n$, respectively. $c$ is the lattice constant along the $z$ direction. $\alpha$, $\beta$, $\gamma$ and $\delta$ are all interband hopping constants. We note that, to describe the band dispersions along the $\Gamma-Z$ direction well, this Hamiltonian has been written as a combination of an in-plane $k\cdot p$ model and a $z$ direction tight-binding model.

The full Hamiltonian with SOC takes the form $H=H_0\otimes \mathbf{1}_2+H_{soc}$ under the spinful bases
$(|1\rangle,|2\rangle,|3\rangle,|4\rangle)\otimes(|\uparrow\rangle,|\downarrow\rangle)$. The symmetry allowed $H_{soc}$ is given by

{\fontsize{9.5}{9.5}\selectfont
\begin{equation}\tag{Eq.S4}
{\fontsize{3}{3.5}\selectfont H_{soc}(\mathbf{k})}=\left(\begin{array}{cccccccc}0&0&0&0&0&i\sqrt{2}\lambda_2&0&0\\0&-\lambda_1&0&0&0&0&0&\sqrt{2}\lambda_3\sin(ck_z)\\0&0&\lambda_1 &0&-i\sqrt{2}\lambda_2&0&0&0\\0&0&0&0&0&0&\sqrt{2}\lambda_3\sin(ck_z)&0\\ 0&0&i\sqrt{2}\lambda_2&0&0&0&0&0\\ -i\sqrt{2}\lambda_2&0&0&0&0&\lambda_1&0&0\\0&0&0&\sqrt{2}\lambda_3\sin(ck_z)&0&0&-\lambda_1&0\\ 0&\sqrt{2}\lambda_3\sin(ck_z)&0&0&0&0&0&0\\\end{array}\right)
\end{equation}
}where $\lambda_1$ and $\lambda_2$ are the on-site out-of-plane and in-plane SOC strength, respectively, while $\lambda_3$ is the first order SOC linear to $k_z$ in the vicinity of the $\Gamma$ point and $Z$ point. Here, we have only kept the $k_z$ dependent linear SOC term ($\lambda_3$), since it is crucial to opening the topologically nontrivial band gap in the $\Gamma-Z$ direction as shown in Fig. 1(d) in the main text.

The fitting parameters of the 8-band effective model are given in Table \ref{table:FST-G} in the below.
\begin{table}[htp]
\caption{Model parameters at the $\Gamma$ point and $Z$ point for FST. For the $Z$ point,
only the adjusted parameters are listed, while the other unlisted parameters are the same as those at the $\Gamma$ point.}\label{table:FST-G}
\begin{center}
{\fontsize{9.5}{9.5}\selectfont
\begin{tabular}{ c|c|c|c|c|c|c|c|c}
\hline
\hline
\multicolumn{9}{c}{$\Gamma$ point}        \\
\hline
$E_1/eV$ & $E_2/eV$ &$E_4/eV$ & $m_{1x}/eV^{-1}\cdot\AA^{-2}$ & $m_{2x}/eV^{-1}\cdot\AA^{-2}$  & $m_{4x}/eV^{-1}\cdot\AA^{-2}$ & $t_{1z}/eV$ & $t_{2z}/eV$  &$t_{4z}/eV$ \\
\hline
0.226   & 0.120  &  0.275  & -0.271  &  -0.151       &  0.131 & -0.004 & 0.076 & 0.426  \\
\hline
$\alpha (eV\cdot\AA^2)$  & $\beta (eV\cdot\AA^2)$ & $\gamma (eV\cdot\AA)$ & $\delta (eV\cdot\AA)$ & $\lambda_1/eV$ &$\lambda_2/eV$  &$\lambda_3/eV$ & $\Delta_1/meV$ & $\Delta_2/meV$\\
\hline
3.048  &  1.524     &   0.003   &   2.448 &  0.050   &  0.025 & 0.008 & 2.5 & 1.7\\
\hline
\multicolumn{9}{c}{$Z$ point}        \\
\hline
$E_1/eV$ &$E_2/eV$ &$E_4/eV$ &$m_{2x}/eV^{-1}\cdot\AA^{-2}$ &$\alpha (eV\cdot\AA^2)$  & $\beta (eV\cdot\AA^2)$  & $\delta (eV\cdot\AA)$ & $\lambda_1/eV$ &$\lambda_2/eV$   \\
\hline
0.225   & 0.291  &  -0.526  & -0.131  &  1.524       &  0.762  & 1.152 & 0.062 & 0.031  \\
\hline
\hline
\end{tabular}
}
\end{center}
\end{table}

When the superconductivity (SC) pairing is introduced, the BdG Hamiltonian of the system is
\begin{equation}\tag{Eq.S5}
H_{BdG}(\mathbf{k},\mathbf{r})=\left(\begin{array}{cc}
H_0(\mathbf{k})\otimes\mathbf{1}_2+H_{soc}(\mathbf{k})-\mu & \Delta_s e^{-i\theta}\tanh(r/\xi) \\ \Delta_s e^{i\theta}\tanh(r/\xi) & -H_0(\mathbf{k})\otimes\mathbf{1}_2-H_{soc}(\mathbf{k})+\mu
\end{array}\right)\ ,
\end{equation}
where we rewrite $\mathbf{k}=-i\nabla$ in the real space , $H_0(\mathbf{k})$ and $H_{soc}(\mathbf{k})$ are given above, and
$\Delta_s=\text{diag}(\Delta_{1},\Delta_{2},\Delta_{2},\Delta_{1})\otimes\mathbf{1}_2$ as defined in the main text. The full Nambu basis here is
{\fontsize{9}{12}\selectfont
\begin{equation}\tag{Eq.S6}
 (a_{1,\mathbf{k}\uparrow},a_{2,\mathbf{k}\uparrow},a_{3,\mathbf{k}\uparrow},a_{4,\mathbf{k}\uparrow}, a_{1,\mathbf{k}\downarrow},a_{2,\mathbf{k}\downarrow},a_{3,\mathbf{k}\downarrow},a_{4,\mathbf{k}\downarrow}, -a^\dag_{1,-\mathbf{k}\downarrow},-a^\dag_{3,-\mathbf{k}\downarrow},-a^\dag_{2,-\mathbf{k}\downarrow},-a^\dag_{4,-\mathbf{k}\downarrow}, a^\dag_{1,-\mathbf{k}\uparrow},a^\dag_{3,-\mathbf{k}\uparrow},a^\dag_{2,-\mathbf{k}\uparrow},a^\dag_{4,-\mathbf{k}\uparrow})
\end{equation}
}where $a_{i,\mathbf{k}\sigma}$ and $a^\dag_{i,\mathbf{k}\sigma}$ are the annihilation and creation operators of fermions in band $i$ with momentum $\mathbf{k}$ and spin $\sigma=\uparrow,\downarrow$, respectively.

We note that $H_{BdG}$  here is an ordinary s-wave SC model in the bulk, and the bulk of FST is indeed not topological~\cite{Miao2012,Yin2015}. However, the presence of the Dirac type surface states allows for the emergence of a 2-dimensional (2D) topological superconductivity (TSC) on the (001) surface. It is our key result that 2D TSC can be realized on the surface of a 3-dimensional normal superconductor (NSC) hosting a topological nontrivial band structure at the Fermi level, which present a realistic and easy way to generate TSC and Majorana Fermions.

\section{SU(2) Berry phase calculations}

\begin{figure}[!h]
\includegraphics[clip,scale=0.9,angle=0]{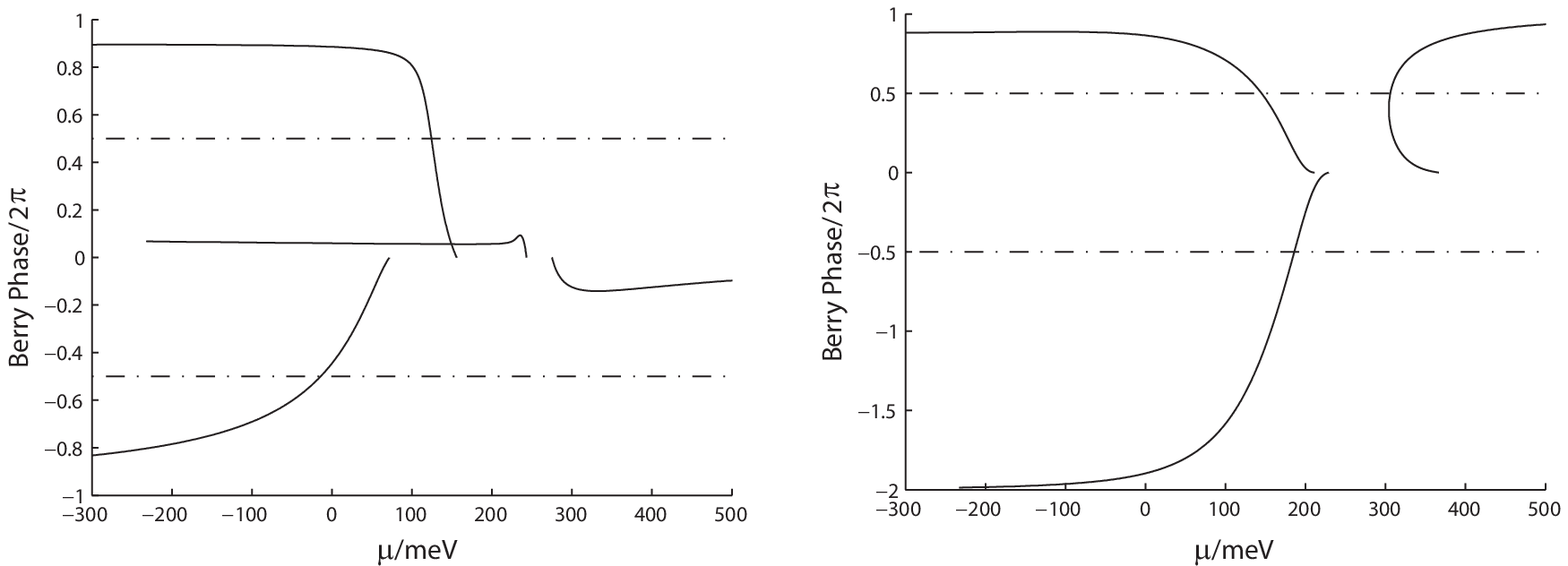}
\caption{The fermi surface SU(2) Berry phase calculated at different chemical potentials $\mu$ in the left panel for the $k_z=0$ (encircling the $\Gamma$ point) plane and the right panel for the $k_z=\pi$ plane (encircling the $Z$ point), respectively. Different solid lines are for different bands, and the dashed-dotted lines are the positions where Berry phase equals $\pm\pi$. Compared to our direct calculations shown in the main text, this criteria gives rather inaccurate predictions.
}
\end{figure}

\begin{figure}[hb]
\includegraphics[clip,scale=0.75,angle=0]{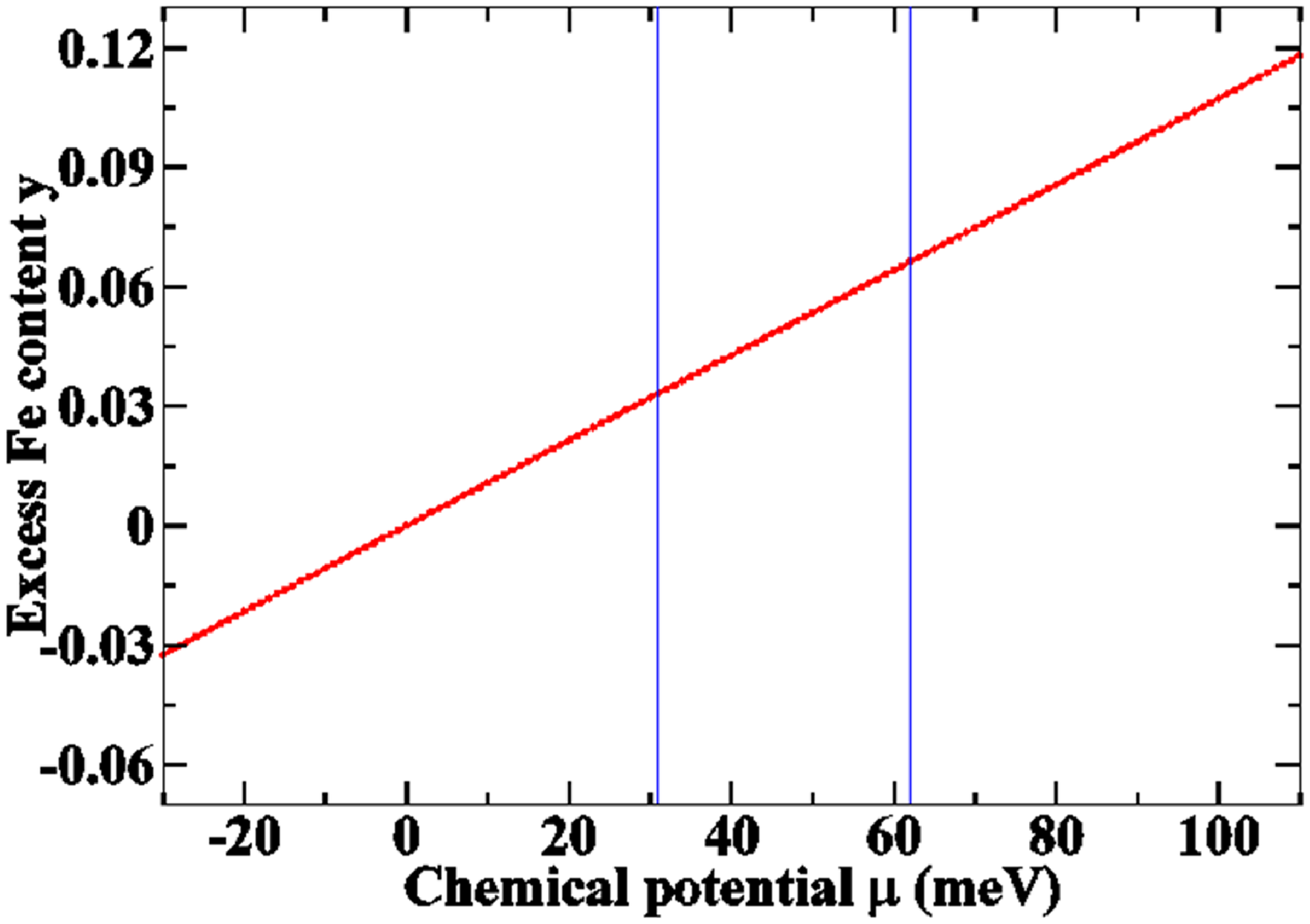}
\caption{(Color online) The chemical potential $\mu$ varying as a function of the excess Fe content $\text{y}$, where the regime between two blue lines corresponds to the TSC phase with $0.03<\text{y}<0.06$.
}
\end{figure}

The authors in Ref. \cite{Hosur2011} argues that the surface TSC/NSC transition point in the zero-pairing limit is determined by the following way: Taking the (001) surface for example, one can calculate the SU(2) Berry phase of the closed lines of the bulk fermi surface in the $k_z=0$ plane (encircling the $\Gamma$ point) and $k_z=\pi$ plane (encircling the Z point),and the transition chemical potential is determined by the requirement that the Berry phase equals to $\pi$. This criterion works well in their single fermi surface model. However, for FST where there are multiple fermi surfaces, this argument does not work well. As shown in Fig. S1, we have calculated the fermi surface Berry phases for different bands in the $k_z=0$ and $k_z=\pi$ plane, respectively. The transition points predicted by this method are very different from our direct calculations, especially for the $k_z=\pi$ plane (encircling the Z point) where we don't see any transitions in the direct calculations. This is understandable because the Berry phase criterion is not a topologically protected result but just a qualitative argument. Their idea is basically to think of the fermi surface in the momentum space as a boundary of a real space TSC. Therefore, its validity relies on the assumptions that the fermi surface is very large and the pairing is local in the momentum space, which are not true in the realistic systems. In particular, the pairing function $\Delta(\mathbf{r})$ near a vortex line is very nonlocal in the momentum space, and thus will couple different fermi surfaces together. As a result, this criterion will work worse when there are multiple fermi surfaces. Instead, we calculate the well-defined topological invariant, Zak phase $\theta_Z$~\cite{Hatsugai2006,Budich2013}, to study the TSC/NSC transition in our work, which gives a consistent conclusion with the gap closing point of the energy spectrum as shown in Fig. 2.

\section{Estimation of the doping level to realize the surface TSC in FST}

According to our results, a little of electrons doping is necessary to achieve the TSC region in FST. Fortunately, many FST superconducting samples are usually self-doped by the excess Fe atoms, and the superconductivity in FST can survive for a wide region with $-0.1<\text{y}<0.1$~\cite{fang2008,Li2009d,Bendele2010}, where $\text{y}$ means the content of the excess (vacancy) Fe atoms in Fe$_{1+\text{y}}$Se$_x$Te$_{1-x}$.  In order to estimate the doping level for realizing the surface chiral TSC in FST, we perform the virtual crystal calculations of FeSe$_{0.5}$Te$_{0.5}$ and plot the chemical potential $\mu$ as a function of the excess Fe content $\text{y}$ in Fig. S2. Our calculated results show that the TSC region is corresponds to $0.03<\text{y}<0.06$, an achievable and superconducting range in experiment.

\section{Robustness of the TSC against the impurities}

\begin{figure}[h]
\includegraphics[clip,scale=0.7,angle=0]{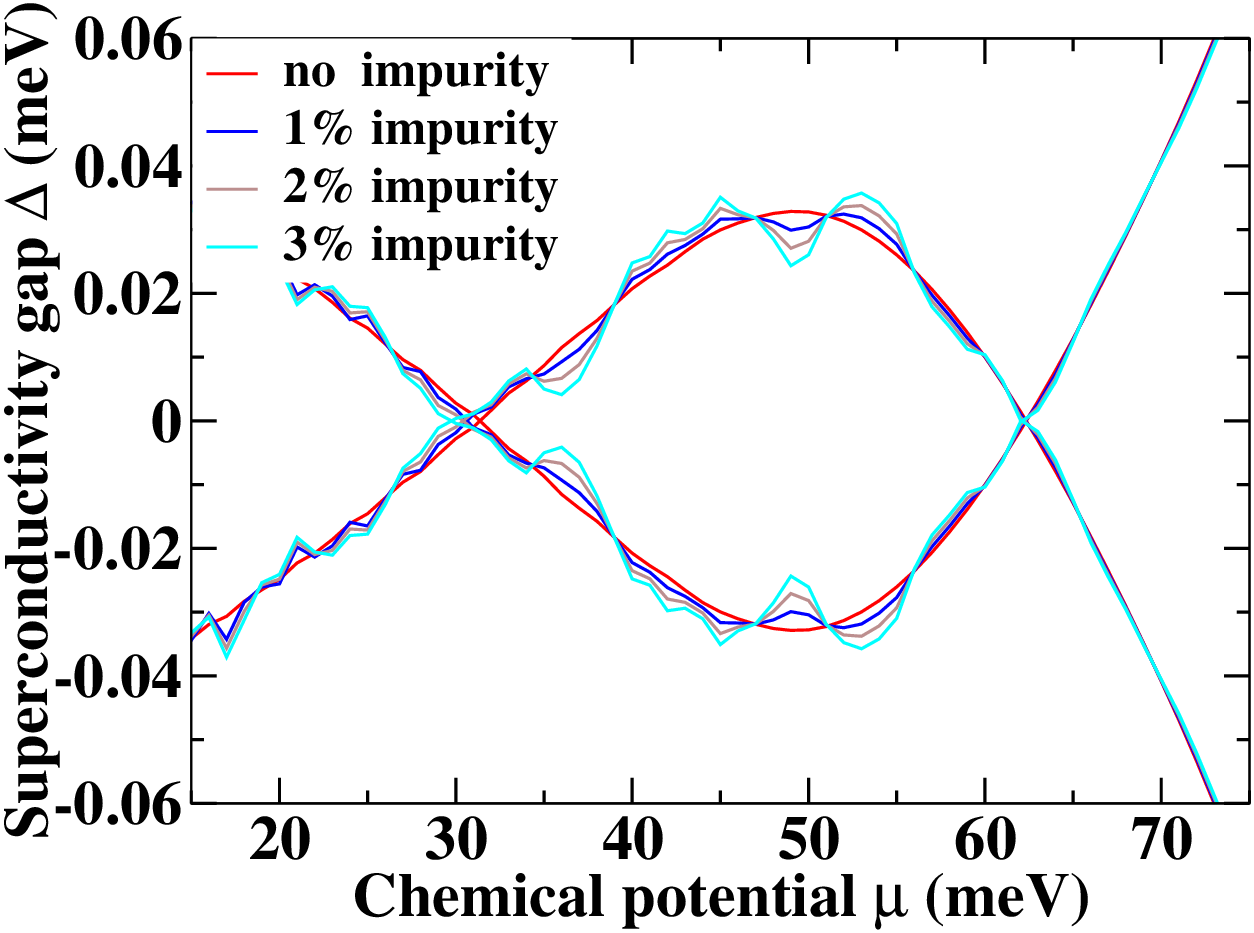}
\caption{(Color online) TSC low energy spectrum against the impurities, in which the results of no impurity, $1\%$ impurities, $2\%$ impurities and $3\%$ impurities are plotted.
}
\end{figure}

In order to study the robustness of the TSC against the weak disorders, we have introduced a random chemical potential disorder with the energy spread $\zeta$ = 0.025 meV, 0.05 meV and 0.075 meV in the vortex line calculation. Physically, such chemical potential disorder is induced by the random impurities. Here, we assume the impurity energy level $\epsilon\ \sim$ 2.5 meV, \emph{i.e.}, the magnitude of the bulk pairing gap $\Delta_{1}$. The energy spread is then estimated using the formula $\zeta = \eta\epsilon$, where $\eta$ is the impurities concentration. The calculated low energy spectrum of the vortex line respect to different impurity level are shown in Fig. S3. As shown in Fig. S3, the low energy spectrum changes little even if the impurity concentration reached up to $3\%$ in the system, which demonstrates that TSC phase is very robust against the impurities and disorders.